\documentclass[aps,prb,fleqn,12pt]{revtex4}
\usepackage{epsfig,color}
\usepackage{graphicx}
\usepackage{pdfpages}
\usepackage{braket}
\usepackage{amssymb,amsmath}
\usepackage{hyperref}
\begin{document}
\title{Coupled spin-orbital fluctuations in a three orbital model for $4d$ and $5d$ oxides with electron fillings $n=3,4,5$ --- Application to {\boldmath $\rm NaOsO_3$}, {\boldmath $\rm Ca_2RuO_4$}, and {\boldmath $\rm Sr_2IrO_4$}}
\author{Shubhajyoti Mohapatra and Avinash Singh}
\email{avinas@iitk.ac.in}
\affiliation{Department of Physics, Indian Institute of Technology, Kanpur - 208016, India}
\date{\today} 
\begin{abstract}
A unified approach is presented for investigating coupled spin-orbital fluctuations within a realistic three-orbital model for strongly spin-orbit coupled systems with electron fillings $n=3,4,5$ in the $t_{2g}$ sector of $d_{yz},d_{xz},d_{xy}$ orbitals. A generalized fluctuation propagator is constructed which is consistent with the generalized self-consistent Hartree-Fock approximation where all Coulomb interaction contributions involving orbital diagonal and off-diagonal spin and charge condensates are included. Besides the low-energy magnon, intermediate-energy orbiton and spin-orbiton, and high-energy spin-orbit exciton modes, the generalized spectral function also shows other high-energy excitations such as the Hund's coupling induced gapped magnon modes. We relate the characteristic features of the coupled spin-orbital excitations to the complex magnetic behavior resulting from the interplay between electronic bands, spin-orbit coupling, Coulomb interactions, and structural distortion effects, as realized in the compounds $\rm NaOsO_3$, $\rm Ca_2RuO_4$, and $\rm Sr_2IrO_4$.
\end{abstract}
%Key words: Spin-orbit Coupling, Generalized Self-Consistent Approach, Coupled Spin-Orbital Excitations, Sodium Osmate, Calcium Ruthenate, Strontium Iridate, Antiferromagnetic Order, Magnetic Anisotropy, Structural Distortions 
%\pacs{75.30.Ds, 71.27.+a, 75.10.Lp, 71.10.Fd}
\maketitle
\newpage

\section{Introduction}

The $4d$ and $5d$ transition metal (TM) oxides exhibit an unprecedented coupling between spin, charge, orbital, and structural degrees of freedom. The complex interplay between the different physical elements such as strong spin-orbit coupling (SOC), Coulomb interactions, and structural distortions results in novel magnetic states and unconventional collective excitations.\cite{krempa_AR_2014,rau_AR_2016,Cao_RPP_2018,zhang_RRL_2018,bertinshaw_AR_2018,zhang_PRB_2020} In particular, the cubic structured $\rm NaOsO_3$ and perovskite structured $\rm Ca_2RuO_4$ and $\rm Sr_2IrO_4$ compounds, corresponding to $d^{\rm n}$ electronic configuration of the TM ion with electron fillings $n$=3,4,5 in the $\rm t_{2g}$ sector, respectively, are at the emerging research frontier as they provide versatile platform for the exploration of SOC-driven phenomena involving collective electronic and magnetic behavior including coupled spin-orbital excitations. 

The different physical elements give rise to a rich variety of nontrivial microscopic features which contribute to the complex interplay. These include spin-orbital-entangled states, band narrowing, spin-orbit gap, and explicit spin-rotation-symmetry breaking (due to SOC), electronic band narrowing due to reduced effective hopping (octahedral tilting and rotation), crystal field induced tetragonal splitting (octahedral compression), orbital mixing (SOC and octahedral tilting, rotation) which self consistently generates induced SOC terms and orbital moment interaction from the Coulomb interaction terms, significantly weaker electron correlation term $U$ compared to $3d$ orbitals and therefore critical contribution of Hund's coupling to local magnetic moment. These microscopic features contribute to the complex interplay in different ways for electron fillings $n$=3,4,5, resulting in significantly different macroscopic properties of the three compounds, which are briefly reviewed below along with experimental observations about the collective and coupled spin-orbital excitations as obtained from recent resonant inelastic X-ray scattering (RIXS) studies. 

The nominally orbitally quenched $d^3$ compound $\rm NaOsO_3$ undergoes a metal-insulator transition (MIT) ($T_{\rm MI} = T_{\rm N}=$ 410 K) that is closely related to the onset of long-range antiferromagnetic (AFM) order.\cite{shi_PRB_2009,calder_PRL_2012,du_PRB_2012,jung_PRB_2013} Various mechanisms, such as Slater-like, magnetic Lifshitz transition, and AFM band insulator have been proposed to explain this unusual and intriguing nature of the MIT.\cite{calder_PRL_2012,vecchio_SREP_2013,kim_PRB_2016,vale_PRL_2018,mohapatra_PRB_2018} Interplay of electronic correlations, Hund's coupling, and octahedral tilting and rotation induced band narrowing near the Fermi level in this weakly correlated compound results in the weakly insulating state with G-type AFM order, with magnetic anisotropy and large magnon gap resulting from interplay of SOC, band structure, and the tetragonal splitting.\cite{mohapatra_PRB_2018,singh_JPCO_2018} The Os $L_3$ resonant edge RIXS measurements at room temperature show four inelastic peak features below 1.5 eV, which have been interpreted to correspond to the strongly gapped ($\sim 58$ meV) dispersive magnon excitations with bandwidth $\sim 100$ meV, excitations (centered at $\sim 1$ eV) within the $t_{\rm 2g}$ manifold, and excitations from $t_{\rm 2g}$ to $e_{\rm g}$ states and ligand-to-metal charge transfer for the remaining two higher-energy peaks.\cite{vale_PRL_2018,calder_PRB_2017,vale_PRB_2018}  The intensity and positions of the three high-energy peaks appear to be essentially temperature independent.

The nominally spin $S$=1 $d^4$ compound $\rm Ca_2RuO_4$ undergoes a MIT at $T_{\rm MI}$=357 K and magnetic transition at $T_{\rm N}$=110 K ($\ll T_{\rm MI}$) via a structural phase transition involving a compressive tetragonal distortion, tilt, and rotation of the $\rm RuO_6$ octahedra.\cite{nakatsuji_JPSP_1997,braden_PRB_1998,alexander_PRB_1999,friedt_PRB_2001} The low-temperature AFM insulating phase is thus characterized by highly distorted octahedra with nominally filled $xy$ orbital and half-filled $yz,xz$ orbitals.\cite{gorelov_PRL_2010,zhang_PRB_2017,sutter_NATCOM_2017} This transition has also been identified in pressure,\cite{nakamura_PRB_2002,steffens_PRB_2005, taniguchi_PRB_2013} chemical substitution,\cite{nakatsuji_PRL_2000,fang_PRB_2001,steffens_PRB_2011} strain,\cite{dietl_APL_2018} and electrical current studies,\cite{nakamura_SREP_2013,okazaki2_JPSJ_2013} and highlights the complex interplay between SOC, Coulomb interactions, and structural distortions. 

Inelastic neutron scattering (INS) \cite{kunkemoller_PRL_2015,jain_NATPHY_2017,kunkemoller_PRB_2017} and Raman \cite{souliou_PRL_2017} studies on $\rm Ca_2RuO_4$ have revealed unconventional low-energy ($\sim$ 50 and 80 meV) excitations interpreted as gapped transverse magnon modes and possibly soft longitudinal (``Higgs-like") or two-magnon excitation modes. From both Ru $L_3$-edge and oxygen $K$-edge RIXS studies, multiple nontrivial excitations within the $t_{\rm 2g}$ manifold were observed recently below 1 eV.\cite{fatuzzo_PRB_2015,das_PRX_2018,gretarsson_PRB_2019} Two low-energy ($\sim$ 80 and 350 meV) and two high-energy ($\sim$ 750 meV and 1 eV) excitations were identified within the limited energy resolution of RIXS. From the incident angle and polarisation dependence of the RIXS spectra, the orbital character of the 80 meV peak was inferred to be mixture of $xy$ and $xz/yz$ states, whereas the 0.4 eV peak was linked to unoccupied $xz/yz$ states. Guided by phenomenological spin models, the low-energy excitations (consisting of multiple branches) were interpreted as composite spin-orbital excitations (also termed as ``spin orbitons").

Finally, SOC induced novel Mott insulating state is realized in the $d^5$ compound $\rm Sr_2IrO_4$,\cite{bjkim_PRL_2008,bjkim_SC_2009} where band narrowing of the spin-orbital-entangled electronic states near the Fermi level plays a critical role in the insulating behavior. The AFM insulating ground state is characterized by the correlation induced insulating gap within the nominally $J$=1/2 bands emerging from the Kramers doublet, which are separated from the bands of the $J$=3/2 quartet by energy $3\lambda/2$, where $\lambda$ is the SOC strength. The RIXS spectra show low-energy dispersive magnon excitations (up to 200 meV), further resolved into two gapped magnon modes with energy gaps $\sim$ 40 meV and 3 meV at the $\Gamma$ point corresponding to out-of-plane and in-plane fluctuation modes, respectively.\cite{jkim1_PRL_2012,igarashi_JPSJ_2014,pincini_PRB_2017,porras_PRB_2019} Weak electron correlation effect and mixing between the $J$=1/2 and 3/2 sectors were identified as contributing significantly to the strong zone-boundary magnon dispersion as measured in RIXS studies.\cite{mohapatra_PRB_2017} In addition, high-energy dispersive spin-orbit exciton modes have also been revealed in RIXS studies in the energy range 0.4-0.8 eV.\cite{kim_NATCOMM_2014} This distinctive mode is also referred to as the spin-orbiton mode,\cite{krupin_JPCM_2016,souri_PRB_2017} and has been attributed to the correlated motion of electron-hole pair excitations across the renormalized spin-orbit gap between the $J$=1/2 and 3/2 bands.\cite{mohapatra_JMMM_2020}

Most of the theoretical studies involving magnetic anisotropy effects and excitations in above systems have mainly focused on phenomenological spin models with different exchange interactions obtained as fitting parameters to the experimental spectra. However, the interpretation of experimental data remains incomplete since the character of the effective spins, the microscopic origin of their interactions, and the microscopic nature of the magnetic excitations are still debated.\cite{rau_AR_2016,bertinshaw_AR_2018,Cao_RPP_2018,zhang_RRL_2018,zhang_PRB_2020} Realistic information about the spin-orbital character of both low and high-energy collective excitations, as inferred from the study of coupled spin-orbital excitations, is clearly important since the spin and orbital degrees of freedom are explicitly coupled, and both are controlled by the different physical elements such as SOC, Coulomb interaction terms, tetragonal compression induced crystal-field splitting between $xy$ and $yz,xz$ orbitals, octahedral tilting and rotation induced orbital mixing hopping terms, and band physics. 

Due to the intimately intertwined roles of the different physical elements, a unified approach is therefore required for the realistic modeling of these systems in which all physical elements are treated on an equal footing. The generalized self-consistent approximation applied recently to the $n=4$ compound $\rm Ca_2RuO_4$ provides such a unified approach.\cite{mohapatra_JPCM_2020} Involving the self-consistent determination of magnetic order within a three-orbital interacting electron model including all orbital-diagonal and off-diagonal spin and charge condensates generated by the different Coulomb interaction terms, this approach explicitly incorporates the complex interplay and accounts for the observed behavior including the tetragonal distortion induced magnetic reorientation transition, orbital moment interaction induced orbital gap, SOC and octahedral tilting induced easy-axis anisotropy, and Coulomb interaction induced anisotropic SOC renormalization. Extension to the $n=5$ compound $\rm Sr_2IrO_4$,\cite{mohapatra_JPCM_2021} provides confirmation of the Hund's coupling induced easy-plane magnetic anisotropy, which is reponsible for the $\sim 40$ meV magnon gap measured for the out-of-plane fluctuation mode.\cite{porras_PRB_2019} 

Towards a generalized non-perturbative formalism unifying the magnetic order and anisotropy effects on one hand and collective excitations on the other, the natural extension of the above generalized condensate approach is therefore to consider the generalized fluctuation propagator in terms of the generalized spin ($\psi_\mu ^\dagger [\sigma^\alpha]\psi_\nu$) and charge ($\psi_\mu ^\dagger [{\bf 1}]\psi_\nu$) operators in the pure spin-orbital basis of the $\rm t_{2g}$ orbitals $\mu,\nu$=$yz,xz,xy$ and spin components $\alpha$=$x,y,z$. The generalized operators include the normal ($\mu=\nu$) spin and charge operators as well as the orbital off-diagonal ($\mu \ne \nu$) cases which are related to the generalized spin-orbit coupling terms ($L_\alpha S_\beta$, where $\alpha,\beta$=$x,y,z$) and the orbital angular momentum operators $L_\alpha$. Constructing the generalized fluctuation propagator as above will ensure that this scheme is fully consistent with the generalized self-consistent approach involving the generalized condensates. 

The different components of the generalized fluctuation propagator will therefore naturally include spin-orbitons and orbitons, corresponding to the spin-orbital ($L_\alpha S_\beta$) and orbital ($L_\alpha$) moment fluctuations, besides the normal spin and charge fluctuations. The normal spin fluctuations will include in-phase and out-of phase fluctuations with respect to different orbitals, the latter being strongly gapped due to Hund's coupling. The spin-orbitons will include the spin-orbit excitons measured in RIXS studies of $\rm Sr_2IrO_4$.

The structure of this paper is as below. The three-orbital model within the $\rm t_{2g}$ sector (including SOC, hopping, Coulomb interaction, and structural distortion terms), and the generalized self-consistent formalism including orbital diagonal and off-diagonal condensates are reviewed in Sec. II and III. After introducing the generalized fluctuation propagator in Sec. IV, results of the calculated fluctuation spectral functions are presented for the cases $n=3,4,5$ (corresponding to the three compounds $\rm NaOsO_3$, $\rm Ca_2RuO_4$, $\rm Sr_2IrO_4$) in Sections V, VI, VII. Finally, conclusions are presented in Sec. VIII. The basis-resolved contributions to the total spectral function showing the detailed spin-orbital character of the collective excitations are presented in the Appendix.

\section{Three orbital model with SOC and Coulomb interactions}
\begin{figure}
\vspace*{0mm}
\hspace*{0mm}
\psfig{figure=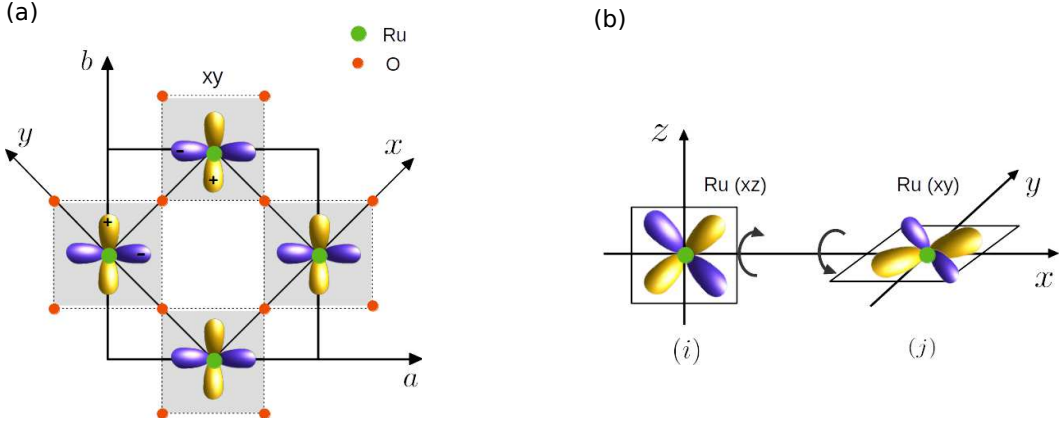,angle=0,width=140mm}
\caption{(a) The common spin-orbital coordinate axes ($x-y$) along the Ru-O-Ru directions, shown along with the crystal axes $a,b$. (b) Octahedral tilting about the crystal $a$ axis is resolved along the $x,y$ axes, resulting in orbital mixing hopping terms between the $xy$ and $yz,xz$ orbitals.} 
\label{axes}
\end{figure}

In the three-orbital ($\mu=yz,xz,xy$), two-spin ($\sigma=\uparrow,\downarrow$) basis defined with respect to a common spin-orbital coordinate axes (Fig. \ref{axes}), we consider the Hamiltonian ${\cal H} = {\cal H}_{\rm band} + {\cal H}_{\rm cf} + {\cal H}_{\rm int} + {\cal H}_{\rm SOC}$ within the $t_{\rm 2g}$ manifold. For the band and crystal field terms together, we consider:
\begin{eqnarray}
{\cal H}_{\rm band+cf} &=& 
\sum_{{\bf k} \sigma s} \psi_{{\bf k} \sigma s}^{\dagger} \left [ \begin{pmatrix}
{\epsilon_{\bf k} ^{yz}}^\prime & 0 & 0 \\
0 & {\epsilon_{\bf k} ^{xz}}^\prime & 0 \\
0 & 0 & {\epsilon_{\bf k} ^{xy}}^\prime + \epsilon_{xy}\end{pmatrix} \delta_{s s^\prime}
+ \begin{pmatrix}
\epsilon_{\bf k}^{yz} & \epsilon_{\bf k}^{yz|xz} & \epsilon_{\bf k}^{yz|xy} \\
-\epsilon_{\bf k} ^{yz|xz} & \epsilon_{\bf k} ^{xz} & \epsilon_{\bf k}^{xz|xy} \\
-\epsilon_{\bf k}^{yz|xy} & -\epsilon_{\bf k}^{xz|xy} & \epsilon_{\bf k} ^{xy} \end{pmatrix} \delta_{\bar{s} s^\prime } \right] \psi_{{\bf k} \sigma s^\prime} \nonumber \\
\label{three_orb_two_sub}
\end{eqnarray} 
in the composite three-orbital, two-sublattice ($s,s'={\rm A,B}$) basis. Here the energy offset $\epsilon_{xy}$ (relative to the degenerate $yz/xz$ orbitals) represents the tetragonal distortion induced crystal field effect. The band dispersion terms in the two groups correspond to hopping terms connecting the same and opposite sublattice(s), and are given by: 
\begin{eqnarray}
\epsilon_{\bf k} ^{xy} &=& -2t_1(\cos{k_x} + \cos{k_y}) \nonumber \\
{\epsilon_{\bf k} ^{xy}} ^{\prime} &=& - 4t_2\cos{k_x}\cos{k_y} - \> 2t_3(\cos{2{k_x}} + \cos{2{k_y}}) \nonumber \\
\epsilon_{\bf k} ^{yz} &=& -2t_5\cos{k_x} -2t_4 \cos{k_y} \nonumber \\
\epsilon_{\bf k} ^{xz} &=& -2t_4\cos{k_x} -2t_5 \cos{k_y}  \nonumber \\
\epsilon_{\bf k} ^{yz|xz} &=&  -2t_{m1}(\cos{k_x} + \cos{k_y}) \nonumber \\
\epsilon_{\bf k} ^{xz|xy} &=&  -2t_{m2}(2\cos{k_x} + \cos{k_y}) \nonumber \\
\epsilon_{\bf k} ^{yz|xy} &=&  -2t_{m3}(\cos{k_x} + 2\cos{k_y}) . 
\label{band}
\end{eqnarray}

Here $t_1$, $t_2$, $t_3$ are respectively the first, second, and third neighbor hopping terms for the $xy$ orbital. For the $yz$ ($xz$) orbital, $t_4$ and $t_5$ are the nearest-neighbor (NN) hopping terms in $y$ $(x)$ and $x$ $(y)$ directions, respectively, corresponding to $\pi$ and $\delta$ orbital overlaps. Octahedral rotation and tilting induced orbital mixings are represented by the NN hopping terms $t_{m1}$ (between $yz$ and $xz$) and $t_{m2},t_{m3}$ (between $xy$ and $xz,yz$). In the $n=4$ case corresponding to the $\rm Ca_2RuO_4$ compound, we have taken hopping parameter values: ($t_1$, $t_2$, $t_3$, $t_4$, $t_5$)=$(-1.0, 0.5, 0, -1.0, 0.2)$, orbital mixing hopping terms: $t_{m1}$=0.2 and $t_{m2}$=$t_{m3}$=0.15 ($\approx 0.2/\sqrt{2}$), and $\epsilon_{xy}=-0.8$, all in units of the realistic hopping energy scale $|t_1|$=150 meV.\cite{khaliullin_PRL_2013,akbari_PRB_2014,feldmaier_arxiv_2019} The choice $t_{m2}=t_{m3}$ corresponds to the octahedral tilting axis oriented along the $\pm(-\hat{x}+\hat{y})$ direction, which is equivalent to the crystal $\mp a$ direction (Fig. \ref{axes}). The $t_{m1}$ and $t_{m2,m3}$ values taken above approximately correspond to octahedral rotation and tilting angles of about $12^\circ$ ($\approx 0.2$ rad) as reported in experimental studies.\cite{steffens_PRB_2005} 

For the on-site Coulomb interaction terms in the $t_{2g}$ basis ($\mu,\nu=yz,xz,xy$), we consider:
\begin{eqnarray}
{\cal H}_{\rm int} &=& U\sum_{i,\mu}{n_{i\mu\uparrow}n_{i\mu\downarrow}} + U^\prime \sum_{i,\mu < \nu,\sigma} {n_{i\mu\sigma} n_{i\nu\overline{\sigma}}} + (U^\prime - J_{\mathrm H}) \sum_{i,\mu < \nu,\sigma}{n_{i\mu\sigma}n_{i\nu\sigma}} \nonumber\\ 
&+& J_{\mathrm H} \sum_{i,\mu \ne \nu} {a_{i \mu \uparrow}^{\dagger}a_{i \nu\downarrow}^{\dagger}a_{i \mu \downarrow} a_{i \nu \uparrow}} + J_{\mathrm P} \sum_{i,\mu \ne \nu} {a_{i \mu \uparrow}^{\dagger} a_{i \mu\downarrow}^{\dagger}a_{i \nu \downarrow} a_{i \nu \uparrow}} \nonumber\\ 
&=& U\sum_{i,\mu}{n_{i\mu\uparrow}n_{i\mu\downarrow}} + U^{\prime \prime}\sum_{i,\mu<\nu} n_{i\mu} n_{i\nu} - 2J_{\mathrm H} \sum_{i,\mu<\nu} {\bf S}_{i\mu}.{\bf S}_{i\nu} 
+J_{\mathrm P} \sum_{i,\mu \ne \nu} a_{i \mu \uparrow}^{\dagger} a_{i \mu\downarrow}^{\dagger}a_{i \nu \downarrow} a_{i \nu \uparrow} 
\label{h_int}
\end{eqnarray} 
including the intra-orbital $(U)$ and inter-orbital $(U')$ density interaction terms, the Hund's coupling term $(J_{\rm H})$, and the pair hopping interaction term $(J_{\rm P})$, with $U^{\prime\prime} \equiv U^\prime-J_{\rm H}/2=U-5J_{\rm H}/2$ from the spherical symmetry condition $U^\prime=U-2J_{\mathrm H}$. Here $a_{i\mu\sigma}^{\dagger}$ and $a_{i\mu \sigma}$ are the electron creation and annihilation operators for site $i$, orbital $\mu$, spin $\sigma=\uparrow ,\downarrow$. The density operator $n_{i\mu\sigma}=a_{i\mu\sigma}^\dagger a_{i\mu\sigma}$, total density operator $n_{i\mu}=n_{i\mu\uparrow}+n_{i\mu\downarrow}=\psi_{i\mu}^\dagger \psi_{i\mu}$, and spin density operator ${\bf S}_{i\mu} = \psi_{i\mu}^\dagger ${\boldmath $\sigma$}$ \psi_{i\mu}$ in terms of the electron field operator $\psi_{i\mu}^\dagger=(a_{i\mu\uparrow}^{\dagger} \; a_{i\mu\downarrow}^{\dagger})$. All interaction terms above are SU(2) invariant and thus possess spin rotation symmetry.  

Finally, for the bare spin-orbit coupling term (for site $i$), we consider the spin-space representation:
\begin{eqnarray} 
{\cal H}_{\rm SOC} (i) & = & -\lambda {\bf L}.{\bf S} = -\lambda (L_z S_z + L_x S_x + L_y S_y) \nonumber \\ 
&=& \left [ \begin{pmatrix} \psi_{yz \uparrow}^\dagger & \psi_{yz \downarrow}^\dagger \end{pmatrix}
\begin{pmatrix} i \sigma_z \lambda /2 \end{pmatrix} 
\begin{pmatrix} \psi_{xz \uparrow} \\ \psi_{xz \downarrow} \end{pmatrix}
+ \begin{pmatrix} \psi_{xz \uparrow}^\dagger & \psi_{xz \downarrow}^\dagger \end{pmatrix}
\begin{pmatrix} i \sigma_x \lambda /2 \end{pmatrix} 
\begin{pmatrix} \psi_{xy \uparrow} \\ \psi_{xy \downarrow} \end{pmatrix} \right . \nonumber \\
& + & \left . \begin{pmatrix} \psi_{xy \uparrow}^\dagger & \psi_{xy \downarrow}^\dagger \end{pmatrix}
\begin{pmatrix} i \sigma_y \lambda /2 \end{pmatrix} 
\begin{pmatrix} \psi_{yz \uparrow} \\ \psi_{yz \downarrow} \end{pmatrix} \right ] + {\rm H.c.}
\label{soc}
\end{eqnarray}
which explicitly breaks SU(2) spin rotation symmetry and therefore generates anisotropic magnetic interactions from its interplay with other Hamiltonian terms. Here we have used the matrix representation:
\begin{equation}
L_z = \begin{pmatrix} 0&-i&0 \\ i&0&0 \\ 0&0&0 \end{pmatrix} ,\;\;
L_x = \begin{pmatrix} 0&0&0 \\ 0&0&-i \\ 0&i&0 \end{pmatrix} ,\;\;
L_y = \begin{pmatrix} 0&0&i \\ 0&0&0 \\ -i&0&0 \end{pmatrix} ,\;\;
\end{equation}
for the orbital angular momentum operators in the three-orbital $(yz,xz,xy)$ basis. 

As the orbital ``hopping" terms in Eq. (\ref{soc}) have the same form as spin-dependent hopping terms $i${\boldmath $\sigma . t'_{ij}$}, carrying out the strong-coupling expansion\cite{hc_JMMM_2019} for the $-\lambda L_z S_z$ term to second order in $\lambda$ yields the anisotropic diagonal (AD) intra-site interactions:
\begin{equation}
[H^{(2)}_{\rm eff}]_{\rm AD}^{(z)}(i) = \frac{4 (\lambda/2)^2 }{U} \left [ S_{yz}^z S_{xz}^z - (S_{yz}^x S_{xz}^x  + S_{yz}^y S_{xz}^y) \right ] 
\label{h_eff}
\end{equation}
between $yz,xz$ moments if these orbitals are nominally half-filled, as in the case of $\rm Ca_2RuO_4$. This term explicitly yields preferential $x-y$ plane ordering (easy-plane anisotropy) for parallel $yz,xz$ moments, as enforced by the relatively stronger Hund's coupling. 

Similarly, from the strong coupling expansion for the other two SOC terms, we obtain additional anisotropic interaction terms which are shown below to yield $C_4$ symmetric easy-axis anisotropy within the easy plane. From the $-\lambda L_x S_x$ and $-\lambda L_y S_y$ terms, we obtain: 
\begin{eqnarray}
[H^{(2)}_{\rm eff}]_{\rm AD}^{(x,y)}(i) &=& \frac{4 (\lambda/2)^2 }{U} \left [S_{xz}^x S_{xy}^x - (S_{xz}^y S_{xy}^y  + S_{xz}^z S_{xy}^z) \right ] \nonumber \\
&+& \frac{4 (\lambda/2)^2 }{U} \left [S_{xy}^y S_{yz}^y - (S_{xy}^x S_{yz}^x  + S_{xy}^z S_{yz}^z) \right ] 
\label{h_eff_2}
\end{eqnarray}
Neglecting the terms involving the $S^z$ components which are suppressed by the easy-plane anisotropy discussed above, we obtain: 
\begin{eqnarray}
[H^{(2)}_{\rm eff}]_{\rm AD}^{(x,y)}(i) &=& -\frac{4 (\lambda/2)^2 }{U} \left [S_{xy}^x (S_{yz}^x - S_{xz}^x) + S_{xy}^y (S_{xz}^y - S_{yz}^y) \right ] \nonumber \\
&=& -\frac{4 (\lambda/2)^2 }{U} f_{xy} S^2 \left [ \sin 2\phi \sin \phi_c \right ] 
\label{h_eff_3}
\end{eqnarray}
where the spin components are expressed as: $S_{xy}^x = f_{xy}S \cos \phi$, $S_{yz}^x = S \cos (\phi-\phi_c)$, $S_{xz}^x = S \cos (\phi+\phi_c)$ (and similarly for the $y$ components) in terms of the overall orientation angle $\phi$ of the magnetic order and the relative canting angle $2\phi_c$ between the $yz,xz$ moments. Here the factor $f_{xy} < 1$ represents the reduced moment for the $xy$ orbital.  

The above expression shows the composite orientation and canting angle dependence of the anisotropic interaction energy having the $C_4$ symmetry. Minimum energy is obtained at orientations $\phi=n\pi/4$ (where $n=1,3,5,7$) since the canting angle has the approximate functional form $\phi_c \approx \phi_c ^{\rm max} \sin 2\phi$ in terms of the orientation $\phi$. Thus, while the easy-plane anisotropy involves only the $yz,xz$ moments, the $xy$ moment plays a crucial role in the easy-axis anisotropy, which is directly relevant for $\rm NaOsO_3$ ($xy$ orbital is also nominally half-filled), but also for $\rm Ca_2RuO_4$ with the factor $f_{xy}$ as incorporated above.

For later reference, we note here that condensates of the orbital off-diagonal (OOD) one-body operators as in Eq. (\ref{soc}) directly yield physical quantities such as orbital magnetic moments and spin-orbital correlations:
\begin{eqnarray}
\langle L_\alpha \rangle & = & -i \left [ \langle \psi_\mu^\dagger \psi_\nu\rangle - \langle \psi_\mu^\dagger \psi_\nu\rangle^* \right ] = 2\ {\rm Im}\langle \psi_\mu^\dagger \psi_\nu\rangle \nonumber \\
\langle L_\alpha S_\alpha \rangle & = & -i \left [ \langle \psi_\mu^\dagger \sigma_\alpha \psi_\nu\rangle - \langle \psi_\mu^\dagger \sigma_\alpha \psi_\nu\rangle^* \right]/2 = {\rm Im}\langle \psi_\mu^\dagger \sigma_\alpha \psi_\nu\rangle\nonumber \\
\lambda^{\rm int}_\alpha & = & (U'' - J_{\rm H}/2)\langle L_\alpha S_\alpha \rangle = (U'' - J_{\rm H}/2) {\rm Im}\langle \psi_\mu^\dagger \sigma_\alpha \psi_\nu\rangle
\label{phys_quan}
\end{eqnarray}
where the orbital pair ($\mu,\nu$) corresponds to the component $\alpha=x,y,z$, and the last equation yields the interaction induced SOC renormalization, as discussed in the next section.

\section{Self-consistent determination of magnetic order}
We consider the various contributions from the Coulomb interaction terms (Eq. \ref{h_int}) in the HF approximation, focussing first on terms with normal (orbital diagonal) spin and charge condensates. The resulting local spin and charge terms can be written as:
\begin{equation}
[{\cal H}_{\rm int}^{\rm HF}]_{\rm normal} = \sum_{i\mu} \psi_{i\mu}^{\dagger} \left [
-\makebox{\boldmath $\sigma . \Delta$}_{i\mu} + {\cal E}_{i\mu} {\bf 1} \right ] \psi_{i\mu} 
\label{h_hf} 
\end{equation}  
where the spin and charge fields are self-consistently determined from:
\begin{eqnarray}
2\Delta_{i\mu}^\alpha &=& U\langle \sigma_{i\mu}^\alpha \rangle + J_{\rm H} \sum_{\nu < \mu} \langle \sigma_{i\nu}^\alpha \rangle \;\;\;\;\;(\alpha=x,y,z) \nonumber \\
{\cal E}_{i\mu} &=& \frac{U\langle n_{i\mu}\rangle}{2} + U'' \sum_{\nu < \mu} \langle n_{i\nu} \rangle 
\label{selfcon}
\end{eqnarray}
in terms of the local charge density $\langle n_{i\mu}\rangle$ and the spin density components $\langle \sigma_{i\mu}^\alpha \rangle$. 

%For $\langle n_{yz}\rangle=\langle n_{xz}\rangle$, the Coulomb renormalized tetragonal splitting is obtained as:
%\begin{eqnarray}
%\tilde{\delta}_{\rm tet} &=& \tilde{\epsilon}_{xz,yz} - \tilde{\epsilon}_{xy} = (\epsilon_{xz,yz} - \epsilon_{xy}) + \left [{\cal E}_{yz,xz} - {\cal E}_{xy} \right ] \nonumber \\ 
%& = & \delta_{\rm tet} + \left [\frac{U\langle n_{yz,xz}\rangle}{2} + U'' \langle n_{yz,xz} + n_{xy}\rangle \right ] - \left [\frac{U\langle n_{xy}\rangle}{2} + 2U'' \langle n_{yz,xz}\rangle \right ] \nonumber \\
%& = & \delta_{\rm tet} + (U'' -U/2)\langle n_{xy} - n_{yz,xz}\rangle 
%\end{eqnarray}
%which shows that the Coulomb renormalization identically vanishes for the realistic relationship $U''=U/2$ for $4d$ orbitals, as discussed in Sec. II. 

There are additional contributions resulting from orbital off-diagonal (OOD) spin and charge condensates which are finite due to orbital mixing induced by SOC and structural distortions (octahedral tilting and rotation). The contributions corresponding to different Coulomb interaction terms are summarized in Appendix A, and can be grouped in analogy with Eq. (\ref{h_hf}) as:
\begin{equation}
[{\cal H}_{\rm int}^{\rm HF}]_{\rm OOD} = \sum_{i,\mu < \nu} \psi_{i\mu}^{\dagger} \left [
-\makebox{\boldmath $\sigma . \Delta$}_{i\mu\nu} + {\cal E}_{i\mu\nu} {\bf 1} \right ] \psi_{i\nu} + {\rm H.c.}
\label{h_hf_od} 
\end{equation}  
where the orbital off-diagonal spin and charge fields are self-consistently determined from:
\begin{eqnarray}
\makebox{\boldmath $\Delta$}_{i\mu\nu} &=& \left (\frac{U''}{2} + \frac{J_{\rm H}}{4} \right ) \langle \makebox{\boldmath $\sigma$}_{i\nu\mu} \rangle + \left (\frac{J_{\rm P}}{2} \right ) \langle \makebox{\boldmath $\sigma$}_{i\mu\nu} \rangle \nonumber \\
{\cal E}_{i\mu\nu} &=& \left (-\frac{U''}{2} + \frac{3J_{\rm H}}{4} \right ) \langle n_{i\nu\mu} \rangle + \left (\frac{J_{\rm P}}{2}\right ) \langle n_{i\mu\nu} \rangle  
\label{sc_od}
\end{eqnarray}
in terms of the corresponding condensates $\langle \makebox{\boldmath $\sigma$}_{i\mu\nu}\rangle \equiv \langle \psi_{i\mu}^{\dagger} \makebox{\boldmath $\sigma$} \psi_{i\nu} \rangle$ and $\langle n_{i\mu\nu} \rangle \equiv \langle \psi_{i\mu}^{\dagger} {\bf 1} \psi_{i\nu} \rangle$. 

The spin and charge condensates in Eqs. \ref{selfcon} and \ref{sc_od} are evaluated using the eigenfunctions ($\phi_{\bf k}$) and eigenvalues ($E_{\bf k}$) of the full Hamiltonian in the given basis including the interaction contributions $[{\cal H}_{\rm int}^{\rm HF}]$ (Eqs. \ref{h_hf} and \ref{h_hf_od}) using:
\begin{equation}
\langle \sigma^\alpha _{i\mu\nu} \rangle \equiv 
\langle \psi_{i\mu} ^\dagger \sigma^\alpha \psi_{i\nu} \rangle =
\sum_{\bf k}^{E_{\bf k} < E_{\rm F} } (\phi_{{\bf k}\mu s \uparrow}^* \;  \phi_{{\bf k}\mu s \downarrow}^* ) [\sigma^\alpha] \left ( \begin{array}{c} \phi_{{\bf k}\nu s\uparrow} \\ \phi_{{\bf k}\nu s\downarrow} \end{array} \right ) 
\end{equation}
for site $i$ on the $s=A/B$ sublattice, and similarly for the charge condensates $\langle n_{i\mu\nu} \rangle \equiv \langle \psi_{i\mu}^{\dagger} {\bf 1} \psi_{i\nu} \rangle$ with the Pauli matrices $[\sigma^\alpha]$ replaced by the unit matrix $[{\bf 1}]$. The normal spin and charge condensates correspond to $\nu=\mu$. For each orbital pair ($\mu,\nu$) = ($yz,xz$), ($xz,xy$), ($xy,yz$), there are three components ($\alpha=x,y,z$) for the spin condensates $\langle \psi_\mu^\dagger \sigma_\alpha \psi_\nu \rangle$ and one charge condensate $\langle \psi_\mu^\dagger {\bf 1} \psi_\nu \rangle$. This is analogous to the three-plus-one normal spin and charge condensates for each of the three orbitals $\mu=yz,xz,xy$. 

The above additional terms involving orbital off-diagonal condensates contribute to orbital physics. Thus, the charge terms lead to coupling of orbital angular momentum operators to weak orbital fields, the spin terms result in interaction-induced SOC renormalization as given in Eq. (\ref{phys_quan}), and the self consistently determined renormalized SOC values are obtained as: 
\begin{equation}
\lambda_\alpha = \lambda + \lambda_\alpha^{\rm int} 
\end{equation}
\label{ren_soc}
for the three components $\alpha=x,y,z$. Results of the self consistent determination of magnetic order  including all orbital diagonal and off-diagonal spin and charge condensates have been presented for the $\rm 4d^4$ compound $\rm Ca_2RuO_4$ recently,\cite{mohapatra_JPCM_2020} illustrating the rich interplay between different physical elements. 

\section{Generalized fluctuation propagator}
Since all generalized spin $\langle \psi_\mu ^\dagger \makebox{\boldmath $\sigma$}\psi_\nu \rangle$ and charge $\langle \psi_\mu ^\dagger \psi_\nu \rangle$ condensates were included in the self consistent determination of magnetic order, the fluctuation propagator must also be defined in terms of the generalized operators. We therefore consider the time-ordered generalized fluctuation propagator:
\begin{equation}
[\chi({\bf q},\omega)] = \int dt \sum_i e^{i\omega(t-t')} 
e^{-i{\bf q}.({\bf r}_i -{\bf r}_j)} 
\times \langle \Psi_0 | T [\sigma_{\mu\nu}^\alpha (i,t) \sigma_{\mu'\nu'}^{\alpha'} (j,t')] |\Psi_0 \rangle 
\end{equation}
in the self-consistent AFM ground state $|\Psi_0 \rangle$, where the generalized spin-charge operators at lattice sites $i,j$ are defined as $\sigma_{\mu\nu}^\alpha = \psi_\mu ^\dagger \sigma^\alpha \psi_\nu$, which include both the orbital diagonal ($\mu=\nu$) and off-diagonal ($\mu\ne\nu$) cases, as well as the spin ($\alpha=x,y,z$) and charge ($\alpha=c$) operators, with $\sigma^\alpha$ defined as Pauli matrices for $\alpha=x,y,z$ and unit matrix for $\alpha=c$. 

In the random phase approximation (RPA), the generalized fluctuation propagator is obtained as:
\begin{equation}
[\chi({\bf q},\omega)]_{\rm RPA} = \frac{2[\chi^0({\bf q},\omega)]}{{\bf 1} - [U][\chi^0({\bf q},\omega)]}
\label{rpa}
\end{equation}
in terms of the bare particle-hole propagator $[\chi^0({\bf q},\omega)]$ which is evaluated by integrating out the electronic degrees of freedom:
\begin{equation}
[\chi^0({\bf q},\omega)]_{\mu\nu\alpha s} ^{\mu'\nu'\alpha ' s'} = \frac{1}{2} 
\sum_{\bf k} \left [ \frac{ \langle {\bf k}| \sigma_{\mu\nu}^\alpha | {\bf k}-{\bf q} \rangle_s 
\langle {\bf k}| \sigma_{\mu'\nu'}^{\alpha '} | {\bf k}-{\bf q} \rangle_{s'} ^* } {E_{{\bf k}-{\bf q}}^\oplus - E_{\bf k}^\ominus + \omega -i\eta} + \frac{ \langle {\bf k} | \sigma_{\mu\nu}^\alpha | {\bf k}-{\bf q} \rangle_s \langle {\bf k} | \sigma_{\mu'\nu'}^{\alpha '} | {\bf k}-{\bf q} \rangle_{s'} ^* } {E_{\bf k}^\oplus - E_{{\bf k}-{\bf q}}^\ominus - \omega -i\eta} \right ]
\end{equation}
The matrix elements in the above expression are evaluated using the eigenvectors of the HF Hamiltonian in the self-consistent AFM state:
\begin{equation}
\langle {\bf k} | \sigma_{\mu\nu}^\alpha | {\bf k}-{\bf q}\rangle_s = (\phi_{{\bf k}\mu\uparrow s}^* \;\; \phi_{{\bf k}\mu\downarrow s}^* ) [\sigma^\alpha] \left (
\begin{array}{c} \phi_{{\bf k}-{\bf q}\ \nu\uparrow s} \\ \phi_{{\bf k}-{\bf q}\ \nu\downarrow s} \end{array} \right )
\end{equation}
and the superscripts $\oplus$ ($\ominus$) refer to particle (hole) states above (below) the Fermi energy.  The subscripts $s,s'$ indicate the two (A/B) sublattices. In the composite spin-charge-orbital-sublattice ($\mu\nu\alpha s$) basis, the $[\chi^0({\bf q},\omega)]$ matrix is of order $72\times 72$, and the form of the $[U]$ matrix in the RPA expression (Eq. \ref{rpa}) is given in Appendix B. 

The spectral function of the excitations will be determined from:
\begin{equation}
{\rm A}_{\bf q}(\omega) = \frac{1}{\pi} {\rm Im \; Tr} [\chi({\bf q}, \omega)]_{\rm RPA}
\label{spectral}
\end{equation}
using the RPA expression for $[\chi({\bf q},\omega)]$. When the collective excitation energies lie within the AFM band gap, it is convenient to consider the symmetric form of the denominator in the RPA expression (Eq. \ref{rpa}):
\begin{equation}
[U][\chi^0({\bf q},\omega)][U] - [U]
\label{rpa_spin}
\end{equation}
and in terms of the real eigenvalues $\lambda_{\bf q}(\omega)$ of this Hermitian matrix, the magnon energies $\omega_{\bf q}$ for momentum ${\bf q}$ are determined by solving for the zeroes: 
\begin{equation}
\lambda_{\bf q}(\omega=\omega_{\bf q}) = 0
\label{dispn}
\end{equation}
corresponding to the poles in the propagator. 

Results of the calculated spectral function will be discussed in the subsequent sections for different electron filling cases ($n=3,4,5$) with applications to corresponding $4d$ and $5d$ transition metal compounds. Broadly, our investigation of the generalized fluctuation propagator will provide information about the dominantly spin, orbital, and spin-orbital excitations, as the generalized spin and charge operators $\psi_\mu ^\dagger \sigma^\alpha \psi_\nu$ include spin ($\mu=\nu$, $\alpha=x,y,z$), orbital ($\mu\ne\nu$, $\alpha=c$), and spin-orbital ($\mu\ne\nu$, $\alpha=x,y,z$) cases. Also included will be the high-energy spin-orbit exciton modes involving particle-hole excitations across the renormalized spin-orbit gap between spin-orbital entangled states of different $J$ sectors, as in the $n=5$ case relevant for the $\rm Sr_2IrO_4$ compound.

%On the other hand, in the $n=3$ case relevant for $\rm NaOsO_3$, where the particle-hole excitations are across the AFM band gap, the high-energy exciton modes will correspond to dominantly magnetic excitons. The character of the low-energy magnon mode also ranges from nearly orbital diagonal and dominantly spin fluctuations for $n=3,4$ to strongly spin-orbital entangled fluctuations for $n=5$. Besides the low-energy (in-phase) modes, the magnon excitations will generally include high-energy modes involving out-of-phase spin fluctuations of different orbitals, as obtained for $n=3,4$, which are strongly gapped due to Hund's coupling. This gapped magnon mode is expected to be quenched in the $n=5$ case as there is only one magnetically active pseudo orbital. We will see that this is replaced by an exciton mode corresponding to out-of-phase fluctuations of spin and orbital moments.

\section{$n=3$ --- application to $\rm NaOsO_3$}

The strongly spin-orbit coupled orthorhomic structured $5d^3$ osmium compound $\rm NaOsO_3$, with nominally three electrons in the Os $t_{2g}$ sector, exhibits several novel electronic and magnetic properties. These include a G-type antiferromagnetic (AFM) structure with spins oriented along the $c$ axis, a significantly reduced magnetic moment $\sim 1 \mu_{\rm B}$ as measured from neutron scattering, a continuous metal-insulator transition (MIT) that coincides with the AFM transition ($T_{\rm N} = T_{\rm MIT}$ = 410 K) as seen in neutron and X-ray scattering, and a large magnon gap of 58 meV as seen in resonant inelastic X-ray scattering (RIXS) measurements indicating strong magnetic anisotropy.\cite{calder_PRL_2012,du_PRB_2012,jung_PRB_2013,calder_PRB_2017} 

Two different mechanisms contributing to SOC-induced easy-plane anisotropy and large magnon gap for out-of-plane fluctuation modes were identified for the weakly correlated $5d^3$ compound $\rm NaOsO_3$ in terms of a simplified picture involving only the normal spin and charge densities.\cite{singh_JPCO_2018,mohapatra_PRB_2018} Both essential ingredients  --- (i) small moment disparity between $yz,xz$ and $xy$ orbitals and (ii) spin-charge coupling effect in presence of tetragonal splitting --- are intrinsically present in the considered three-orbital model on the square lattice. A realistic representation of magnetic anisotropy in $\rm NaOsO_3$ is therefore provided by the considered model, while maintaining uniformity of lattice structure across the $n=3,4,5$ cases considered in order to keep the focus on coupled spin-orbital fluctuations. 

The first mechanism involves the SOC-induced anisotropic interaction terms as in Eq. (\ref{h_eff}) resulting from the three SOC terms $-\lambda L_\alpha S_\alpha$ for $\alpha=x,y,z$. Due to the small moment disparity $m_{yz,xz} > m_{xy}$ resulting from the broader $xy$ band, the interaction term in Eq. (\ref{h_eff}) dominates over the other two terms, leading to the easy-plane anisotropy for parallel $yz,xz$ moments enforced by the Hund's coupling. With increasing $U$, this effect weakens as the moments saturate $m_{yz,xz,xy} \approx 1$ in the large $U$ limit. In the second mechanism, the SOC induced decreasing $xy$ orbital density $n_{xy}$ with spin rotation from $z$ direction to $x-y$ plane couples to the tetragonal distortion term, and for positive $\epsilon_{xy}$ the energy is minimized for spin orientation in the $x-y$ plane.  

We will consider the parameter set values $U=4$, $J_{\rm H}=U/5$, $U''=U-5J_{\rm H}/2$, bare SOC value $\lambda$=1.0, and $\epsilon_{\rm xy}=0.5$ unless otherwise indicated, with the hopping energy scale $|t_1|$=300 meV. Thus, $U=1.2$ eV, $\lambda$=0.3 eV, $\epsilon_{\rm xy}=0.15$ eV, which are realistic values for the $\rm Na Os O_3$ compound. Initially, we will also set $t_{m1,m2,m3}=0$ for simplicity, and focus on the easy-plane anisotropy and large magnon gap for out-of-plane fluctuations. 

Self consistent determination of magnetic order using the generalized approach discussed in Sec. III confirms the easy-plane anisotropy. Starting in nearly $z$ direction, the AFM order direction self consistently approaches the $x-y$ plane in a few hundred iterations. Initially, we will discuss magnetic excitations in the self consistent state with AFM order along the $\hat{x}$ or $\hat{y}$ directions. Although these orientations correspond to metastable states as discussed later, they provide convenient test cases for explicitly confirming the gapless in-plane and gapped out-of-plane magnon modes in the generalized fluctuation propagator calculation. 

\begin{figure}
\vspace*{0mm}
\hspace*{0mm}
\psfig{figure=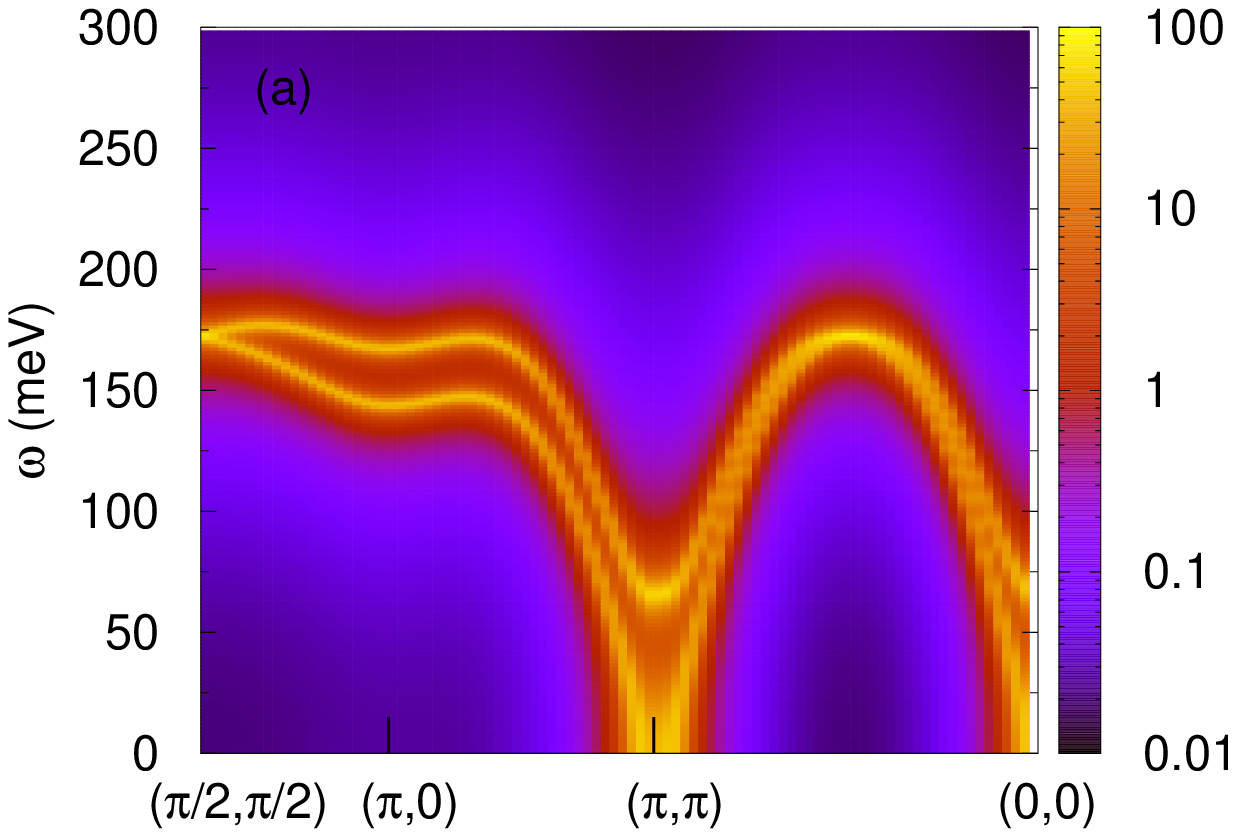,angle=0,width=80mm}
\psfig{figure=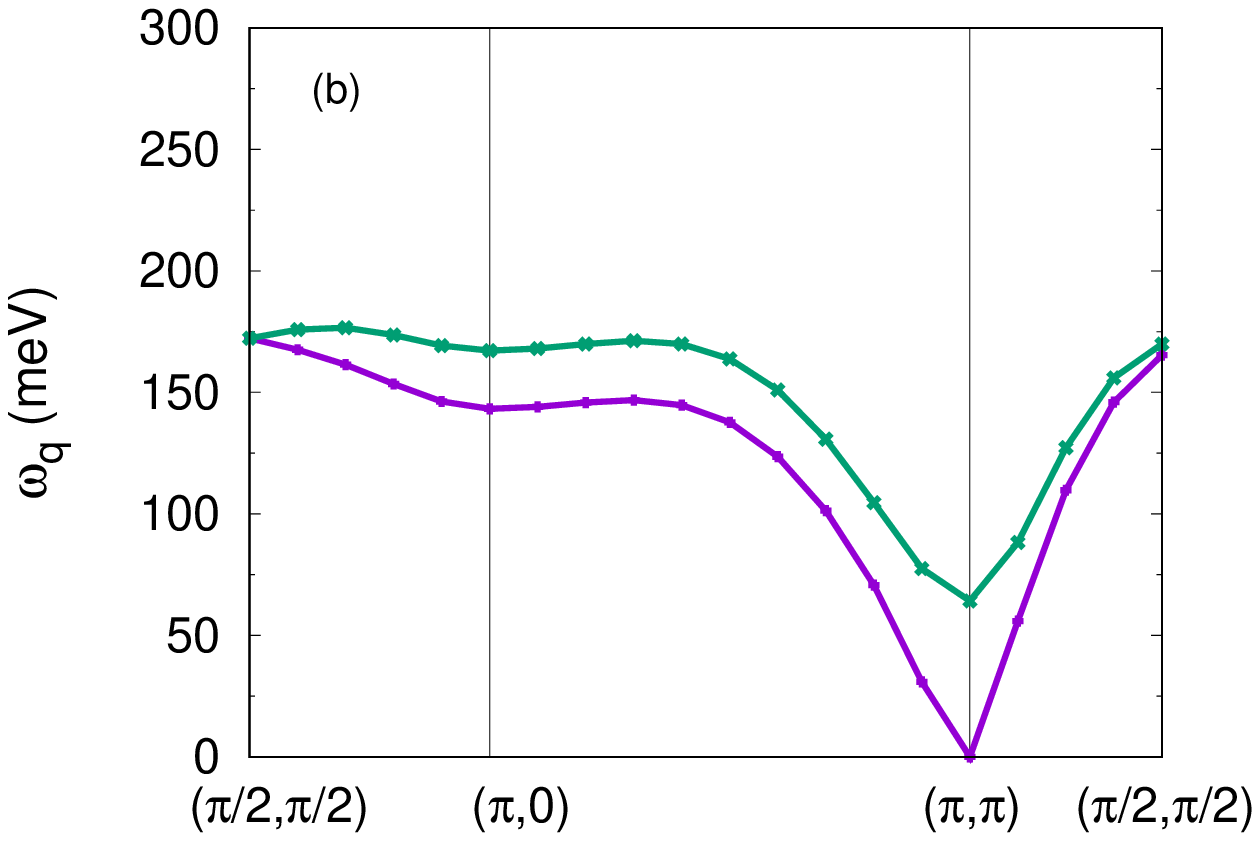,angle=0,width=80mm}
\caption{(a) Low energy part of the calculated spectral function from the generalized fluctuation propagator shows the magnon excitations in the self-consistent state with planar AFM order, and (b) magnon dispersion showing the gapless and gapped modes corresponding to in-plane and out-of-plane fluctuations.} 
\label{spfn1}
\end{figure}

\begin{figure}
\hspace*{0mm}
\psfig{figure=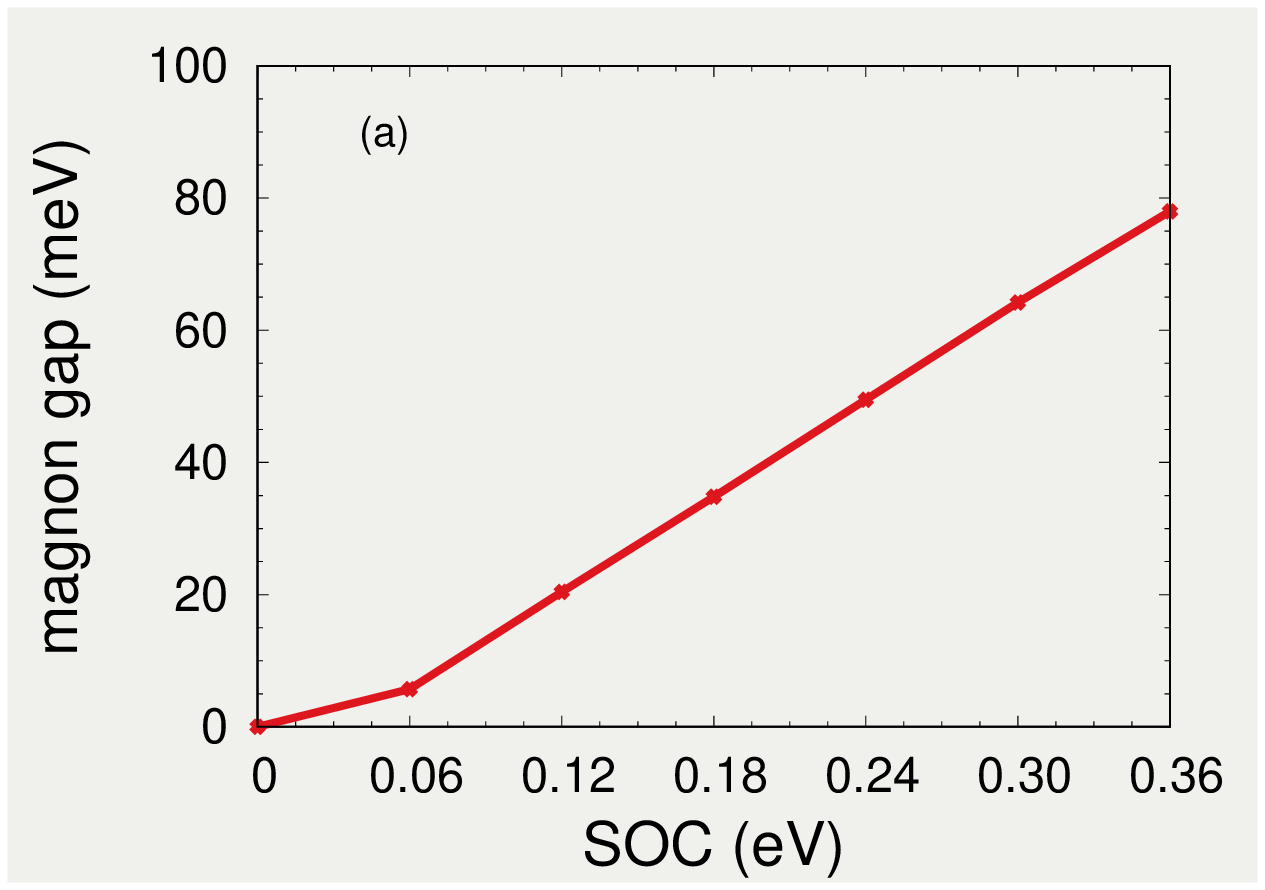,angle=0,width=53mm}
\psfig{figure=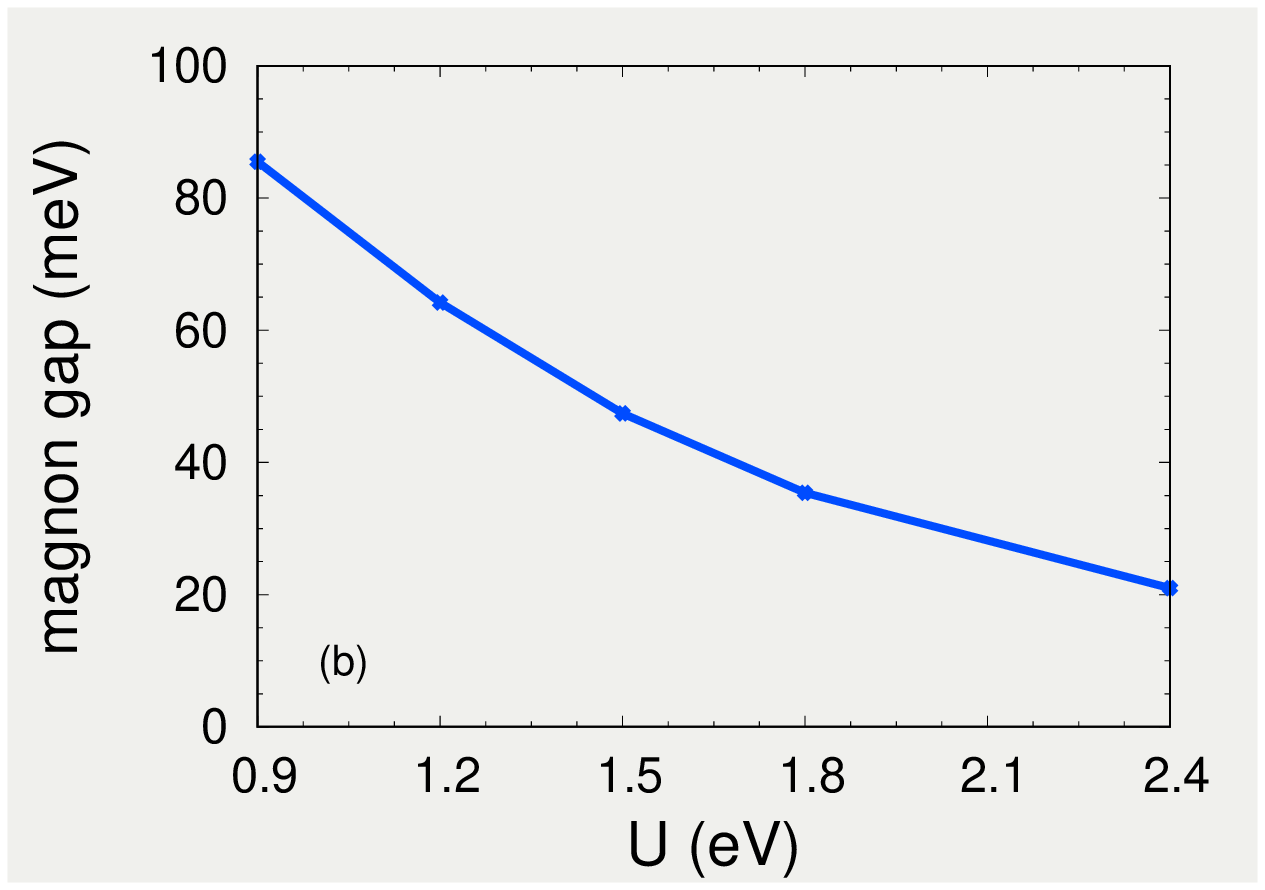,angle=0,width=53mm}
\psfig{figure=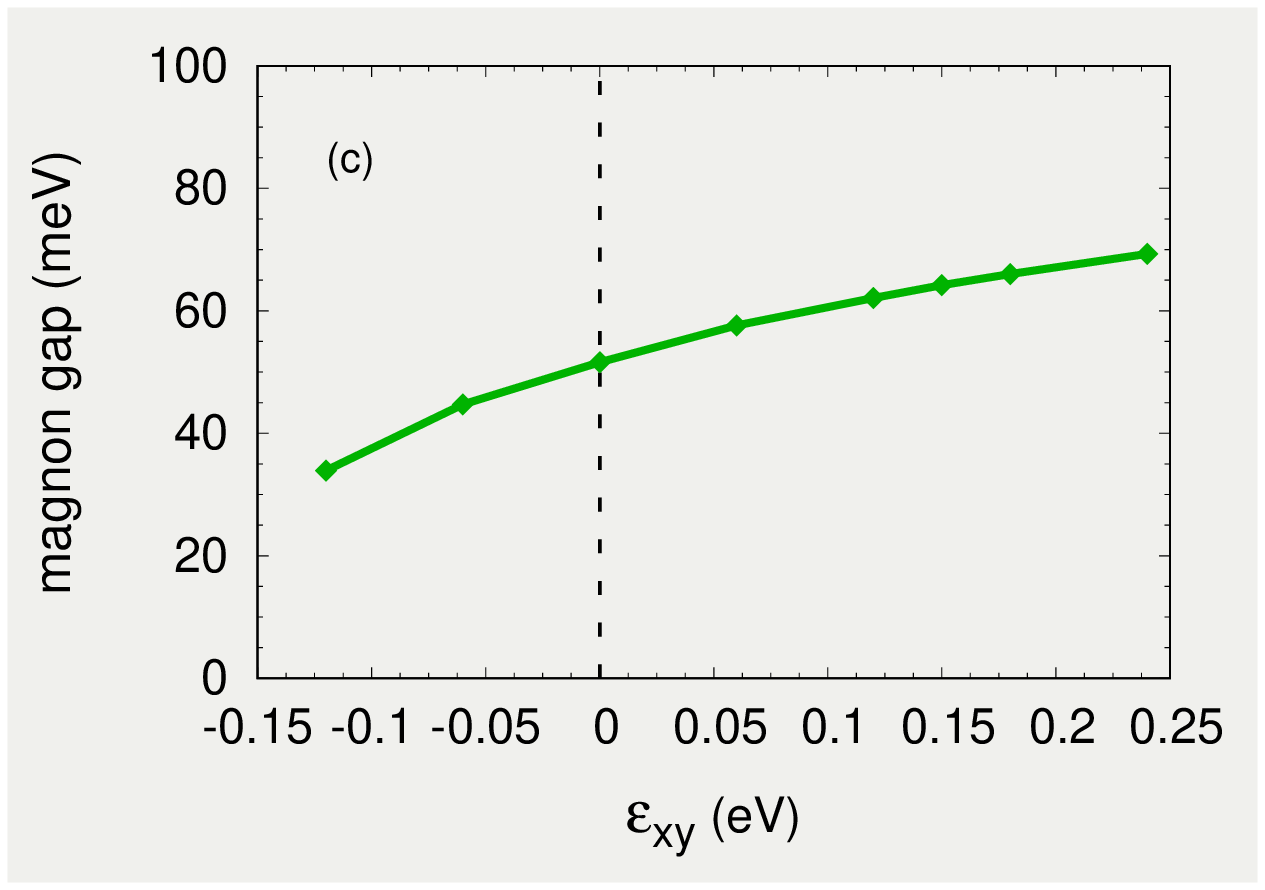,angle=0,width=53mm}
\caption{Variation of the calculated magnon gap showing effects of (a) SOC, (b) Hubbard $U$, and (c) tetragonal distortion $\epsilon_{\rm xy}$, on the easy-plane magnetic anisotropy.} 
\label{wgap1}
\end{figure}

The low-energy part of the calculated spectral function using Eq. (\ref{spectral}) is shown in the Fig. \ref{spfn1}(a) as an intensity plot for ${\bf q}$ along symmetry directions of the Brillouin zone. The gapless and gapped modes corresponding to in-plane and out-of-plane fluctuations reflect the easy-plane magnetic anisotropy. The calculated gap energy 60 meV is close to the measured spin wave gap of 58 meV in $\rm NaOsO_3$. Also shown for comparison in Fig. \ref{spfn1}(b) is the magnon dispersion calculated from the poles of the RPA propagator as described in Sec. IV. Focussing on the magnon gap in Fig. \ref{spfn1}(b), which provides a measure of the SOC induced easy-plane anisotropy, effects of various physical quantities are shown in Fig. \ref{wgap1}.
  
% by incorporating the spin-charge coupling effect as a magnon self energy correction, 

%(c) Compared to the RPA result with only orbital diagonal terms ($\mu\mu$), the significant enhancement in the magnon gap illustrates the importance of including the orbital off-diagonal terms ($\mu\nu$) 

The gapless Goldstone mode corresponding to in-plane rotation of AFM ordering direction in the $x-y$ plane involves only small changes in spin densities $\langle \psi_\mu^\dagger \sigma^\alpha \psi_\mu \rangle$ for $\alpha=x,y$ and $\mu=yz,xz$, and also in generalized spin densities $\langle \psi_\mu^\dagger \sigma^\alpha \psi_\nu \rangle$ for $\alpha=x,y$ and $\mu=yz,xz$ with fixed $\nu=xy$. For example, the magnetization values $m_{yz}^x=0.82$ and $m_{xz}^x=0.84$ change to $m_{yz}^y=0.84$ and $m_{xz}^y=0.82$ when the ordering is rotated from $x$ to $y$ direction. Thus, the Goldstone mode is nearly pure spin mode and the small orbital character reflects the effectively suppressed spin-orbital entangement in the $n=3$ AFM state. In contrast, the $n=5$ case corresponding to $\rm Sr_2IrO_4$ shows strongly coupled spin-orbital character of the Goldstone mode (Appendix C) due to the extreme spin-orbital entanglement.

%These changes, however, are consistent with rotation of the common spin-orbital coordinate system which will leave the spin-orbit coupling term invariant. On the other hand, out-of-plane rotations will involve $yz,xz$ orbitals going into $xy$ orbital, which are not equivalent due to both moment disparity and the tetragonal splitting term $\epsilon_{xy}$, resulting in the finite magnon gap.  

We now consider the easy-axis anisotropy effects in our self consistent determination of magnetic order. With respect to the AFM order orientation (azimuthal angle $\phi$) within the easy ($x-y$) plane, we find an easy-axis anisotropy along the diagonal orientations $\phi = n\pi/4$ ($n=1,3,5,7$) even for no octahedral tilting. This anisotropy is due to the orientation and canting angle dependent anisotropic interaction (Eq. \ref{h_eff_3}) as discussed in Sec. II. The anisotropic interaction energy vanishes for $\phi$ along the $x,y$ axes (hence the gapless in-plane mode in Fig. \ref{spfn1}), and is significant near the diagonal orientations, resulting in easy-axis anisotropy and small relative canting between $yz,xz$ moments which is explicitly confirmed in our self-consistent calculation. 

The resulting $C_4$ symmetry of the easy-axis $\pm(\hat{x}\pm\hat{y}$) is reduced to $C_2$ symmetry $\pm(\hat{x}-\hat{y}$) in the presence of octahedral tilting. The important anisotropy effects of the octahedral tilting induced inter-site DM interactions are discussed below. We find that the DM axis lies along the crystal $b$ axis, leading to easy axis direction along the crystal $a$ axis. Both these directions are interchanged in comparison to the  $\rm Ca_2RuO_4$ case, which follows from a subtle difference in the present $n=3$ case as explained below.

Following the analysis carried out for the $\rm Ca_2RuO_4$ compound,\cite{mohapatra_JPCM_2020} within the usual strong-coupling expansion in terms of the normal ($t$) and spin-dependent ($t_x',t_y'$) hopping terms induced by the combination of SOC and orbital mixing hopping terms $t_{m2,m3}$ due to octahedral tilting, the DM interaction terms generated in the effective spin model are obtained as:
\begin{eqnarray}
[H_{\rm eff}^{(2)}]_{\rm DM}^{(x,y)} &=& \frac{8tt_x'}{U} \sum_{\langle i,j \rangle_x} \hat{x}.({\bf S}_{i,xz} \times {\bf S}_{j,xz}) 
+ \frac{8tt_y'}{U} \sum_{\langle i,j\rangle_y} \hat{y}.({\bf S}_{i,yz} \times {\bf S}_{j,yz}) \nonumber \\ 
& \approx & \frac{8t|t_x'|}{U} \sum_{\langle i,j \rangle} 
(\hat{x} + \hat{y}).({\bf S}_{i,yz} \times {\bf S}_{j,yz}) 
\end{eqnarray}
where we have taken $t_x'=-t_y'=-$ive and $S_{i,xz}^x = S_{i,yz}^x$ (due to Hund's coupling) as earlier, but with $S_{i,xz}^z = - S_{i,yz}^z$ for the $n=3$ case as obtained in our self consistent calculation which is discussed below. The effective DM axis ($\hat{x} + \hat{y}$) is thus along the crystal $b$ axis (Fig. \ref{axes}) for the $yz$ orbital, resulting in easy-axis anisotropy along the crystal $a$ direction, as well as spin canting about the DM axis in the $z$ direction.

\begin{table}[t] \vspace*{0mm}
\caption{Self consistently determined magnetization and density values for the three orbitals ($\mu$) on the two sublattices ($s$), showing easy-axis anisotropy along the crystal $a$ axis due to octahedral tilting induced DM interaction. Here $t_{m2,m3}=0.15$.} 
\centering % centering table
\begin{tabular}{l c c c c} \\  % creating 5 columns 
\hline\hline % inserting double-line
$\mu$ (s) & $m_\mu^x$ & $m_\mu^y$ & $m_\mu^z$ & $n_\mu$ \\ [0.5ex]
\hline % inserts single-line
$yz$ (A) & 0.598 & $-$0.557 & 0.006 & 1.012 \\[1ex]
$xz$ (A) & 0.557 & $-$0.598 & $-$0.006 & 1.012 \\[1ex]
$xy$ (A) & 0.541 & $-$0.541 & 0.0 & 0.977 \\[1ex] \hline
\end{tabular} \hspace{10mm}
\begin{tabular}{l c c c c} \\  % creating 5 columns 
\hline\hline % inserting double-line
$\mu$ (s) & $m_\mu^x$ & $m_\mu^y$ & $m_\mu^z$ & $n_\mu$ \\ [0.5ex]
\hline % inserts single-line
$yz$ (B) & $-$0.598 & 0.557 & 0.006 & 1.012 \\[1ex]
$xz$ (B) & $-$0.557 & 0.598 & $-$0.006 & 1.012 \\[1ex]
$xy$ (B) & $-$0.541 & 0.541 & 0.0 & 0.977 \\[1ex] \hline 
% inserts single-line
\end{tabular}
\label{table1}
\end{table}
\begin{table}
\caption{Self consistently determined renormalized SOC values $\lambda_\alpha = \lambda+\lambda_\alpha^{\rm int}$ and the orbital magnetic moments $\langle L_\alpha \rangle$ for $\alpha=x,y,z$ on the two sublattices. Bare SOC value $\lambda$=1.0.} 
\centering % centering table
\begin{tabular}{l r r r r r r} \\  % creating 6 columns 
\hline\hline % inserting double-line
$s$ & $\lambda_x$ & $\lambda_y$ & $\lambda_z$ & $\langle L_x \rangle$ & $\langle L_y \rangle$ & $\langle L_z \rangle$ \\ [0.5ex]
\hline % inserts single-line
A & 1.179 & 1.179 & 1.364 & 0.032 & $-$0.032 & 0.0 \\[1ex]
B & 1.179 & 1.179 & 1.364 & $-$0.032 & 0.032 & 0.0 \\[1ex]
\hline % inserts single-line
\end{tabular}
\label{table2}
\end{table}

Results for various physical quantities are shown in Tables I and II. Starting with initial orientation along the $\hat{x}$ or $\hat{y}$ directions, the AFM order direction self consistently approaches the easy-axis direction in a few hundred iterations, explicitly exhibiting the strong easy-axis anisotropy within the easy ($x-y$) plane due to the octahedral tilting induced DM interaction, along with small spin canting in the $z$ direction about the DM axis. The small moment disparity $m_{yz,xz}>m_{xy}$ and the negligible orbital moments can also be seen here explicitly. The renormalized SOC strength $\lambda_z$ is enhanced relative to the other two components, which further reduces the SOC induced frustration in this system with nominally one electron in each of the three orbitals. 

\begin{figure}[t]
\vspace*{-0mm}
\hspace*{0mm}
\psfig{figure=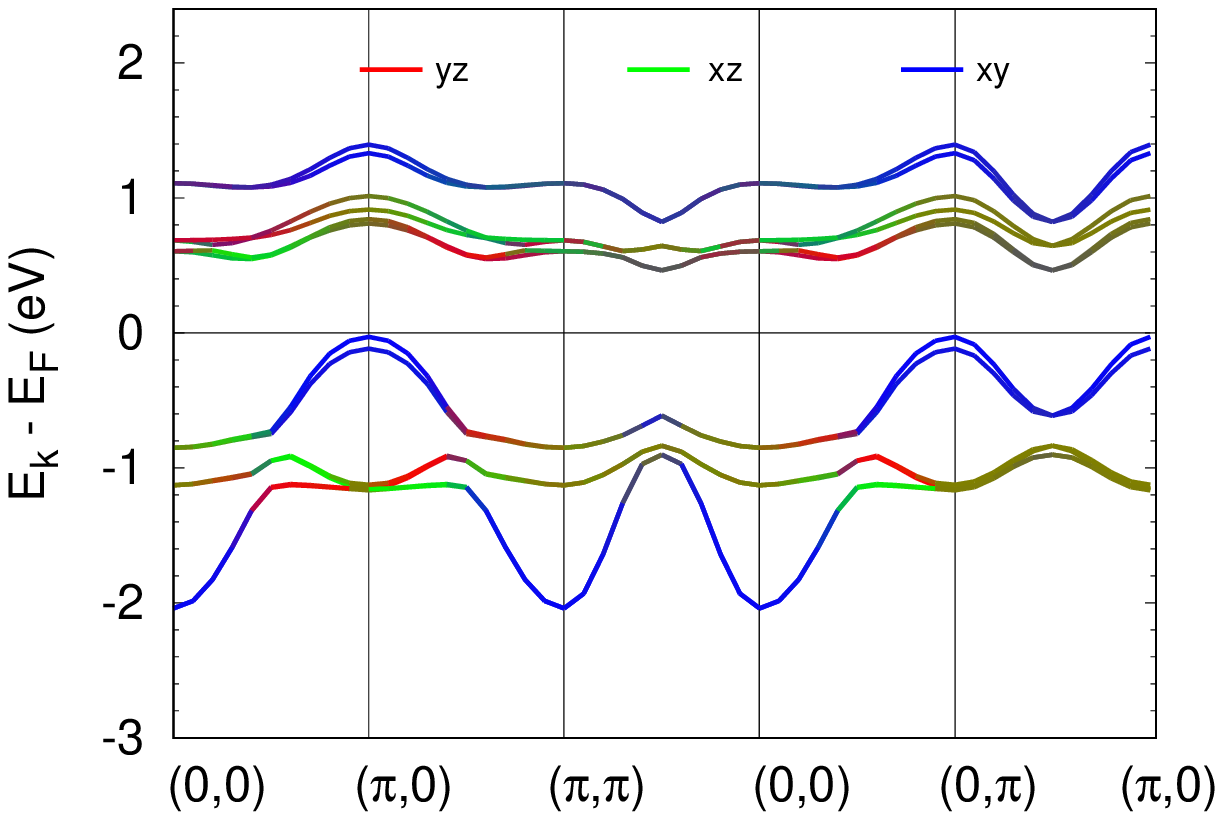,angle=0,width=80mm}
\psfig{figure=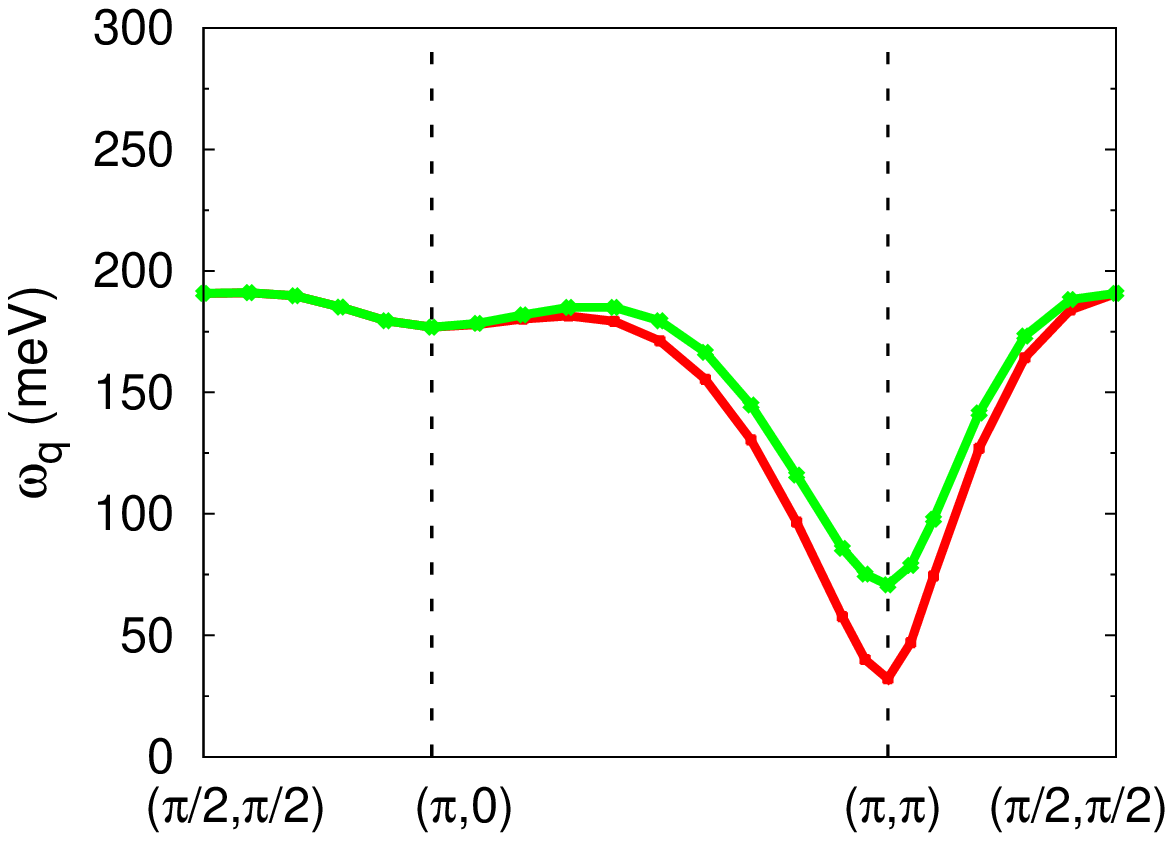,angle=0,width=80mm}
\caption{(a) Calculated orbital resolved electronic band structure in the self-consistent state with AFM order along the crystal $a$ axis due to octahedral tilting induced DM interaction. Here $t_{m2,m3}=0.15$. Colors indicate dominant orbital weight: red ($yz$), green ($xz$), blue ($xy$). (b) Magnon dispersion for the magnetic order as given in Table I, showing that both in-plane and out-of-plane modes are appreciably gapped due to the easy-axis and easy-plane anisotropies.} 
\label{band1}
\end{figure}

With octahedral tilting included, the orbital resolved electronic band structure in the self-consistent AFM state (Fig. \ref{band1}(a)) shows the AFM band gap between valence and conduction bands, SOC induced orbital mixing and band splittings, the fine splitting due to octahedral tilting, and the asymmetric bandwidth for $xy$ orbital bands characteristic of the 2nd neighbor hopping term $t_2$ which connects the same magnetic sublattice. The calculated magnon dispersion evaluated using Eq. \ref{dispn} is shown in Fig. \ref{band1}(b). As expected, both in-plane and out-of-plane magnon modes are gapped due to the easy-axis and easy-plane anisotropies discussed above.  

%\begin{figure}[b]
%\hspace*{0mm}
%\psfig{figure=fig5.eps,angle=0,width=90mm}
%\caption{Magnon dispersion for the self consistently determined magnetic order as given in Table I in the presence of octahedral tilting ($t_{m2,m3}=0.15$), showing that both modes are appreciably gapped due to the easy-plane and easy-axis anisotropies.} 
%\label{magnon}
%\end{figure}

\begin{figure}[t]
\hspace*{0mm}
\psfig{figure=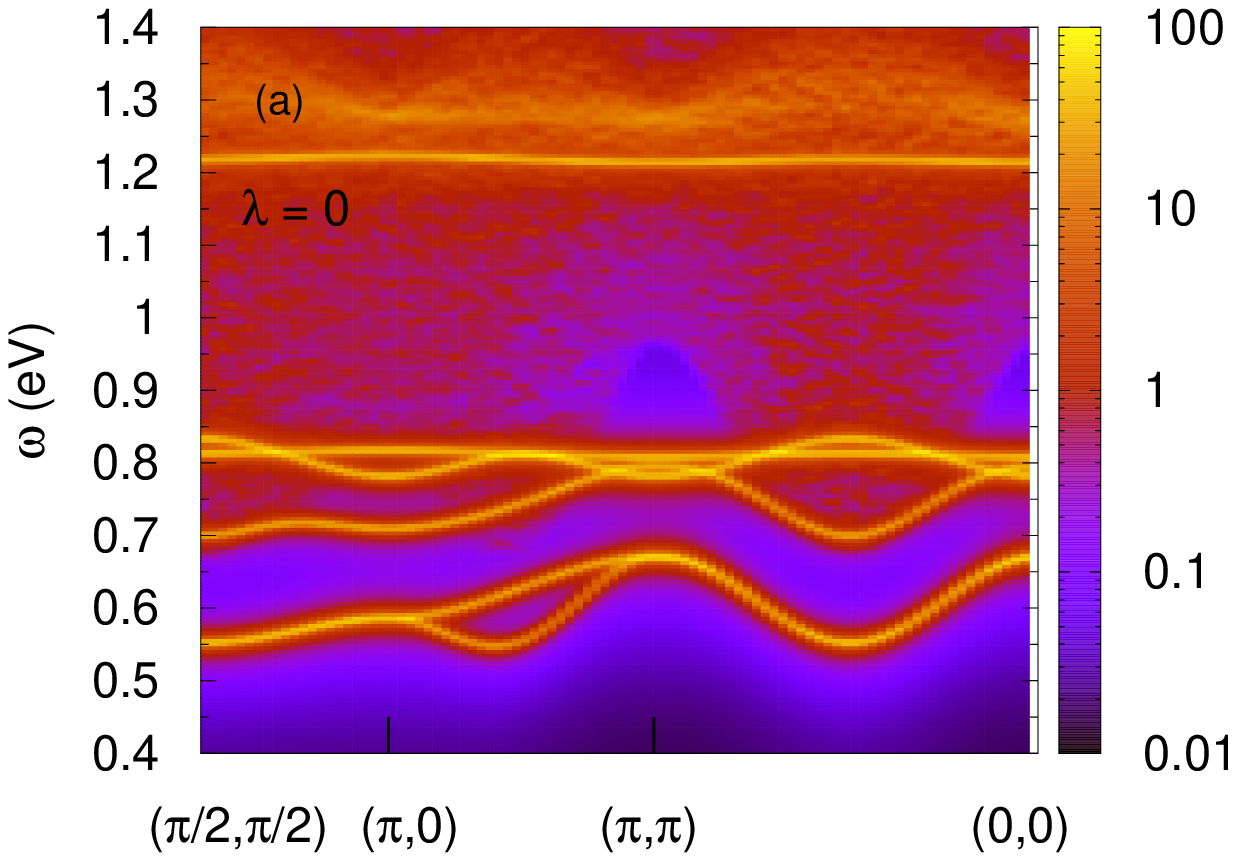,angle=0,width=70mm}
\psfig{figure=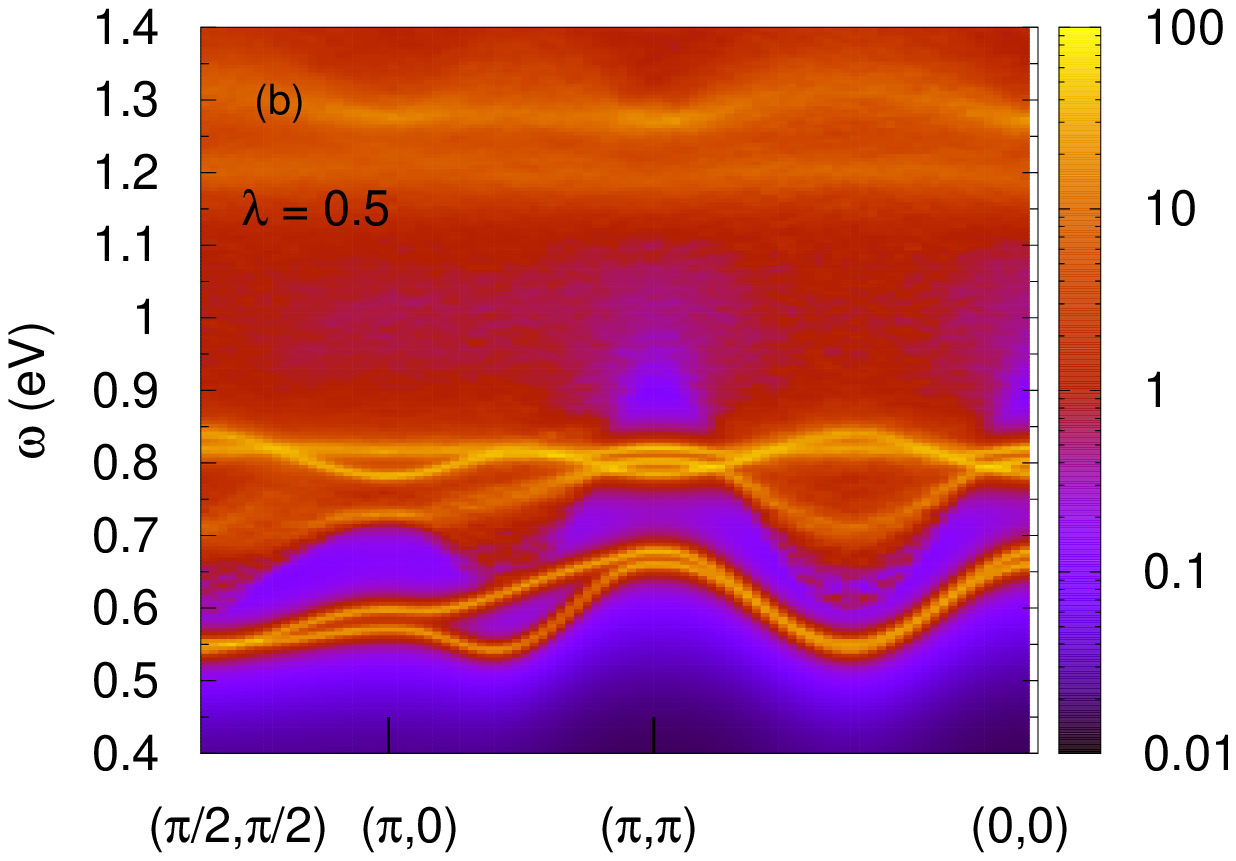,angle=0,width=70mm}
\psfig{figure=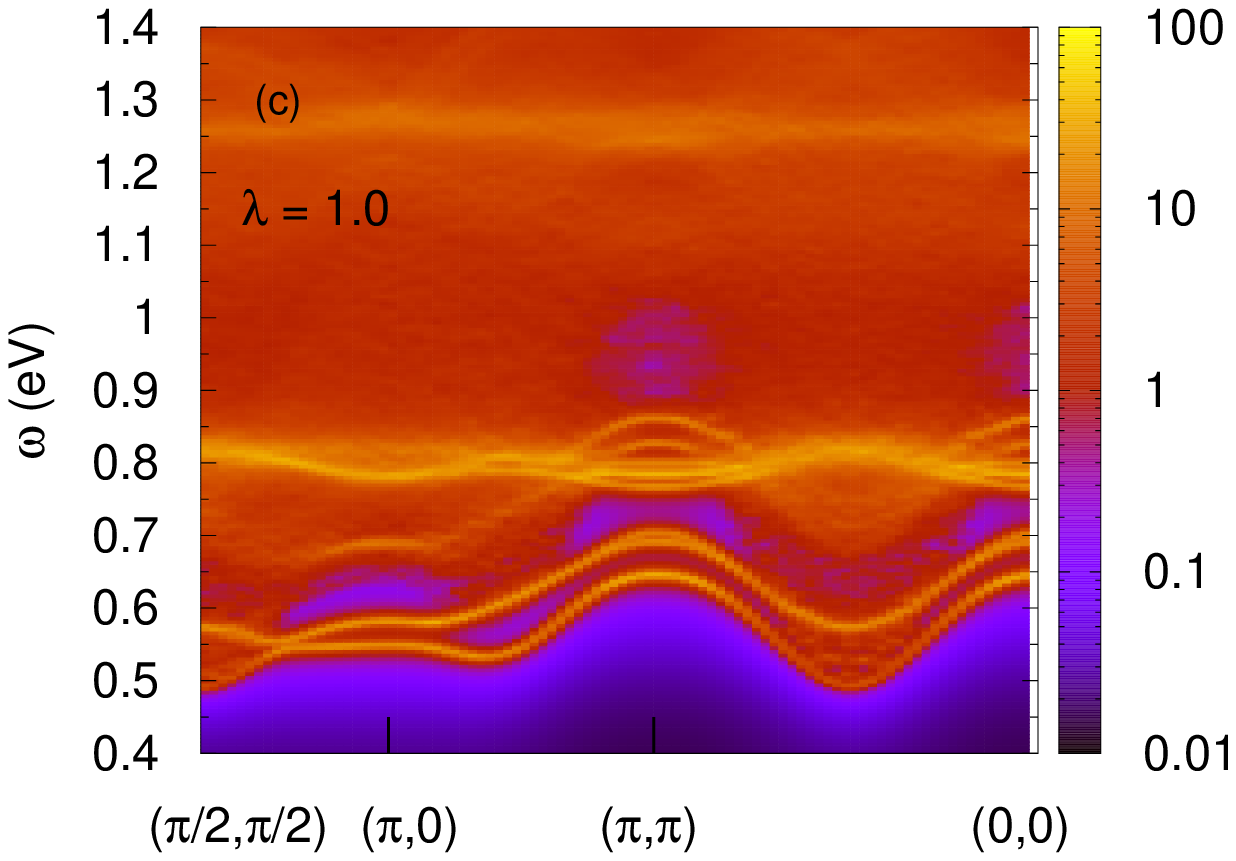,angle=0,width=70mm}
\psfig{figure=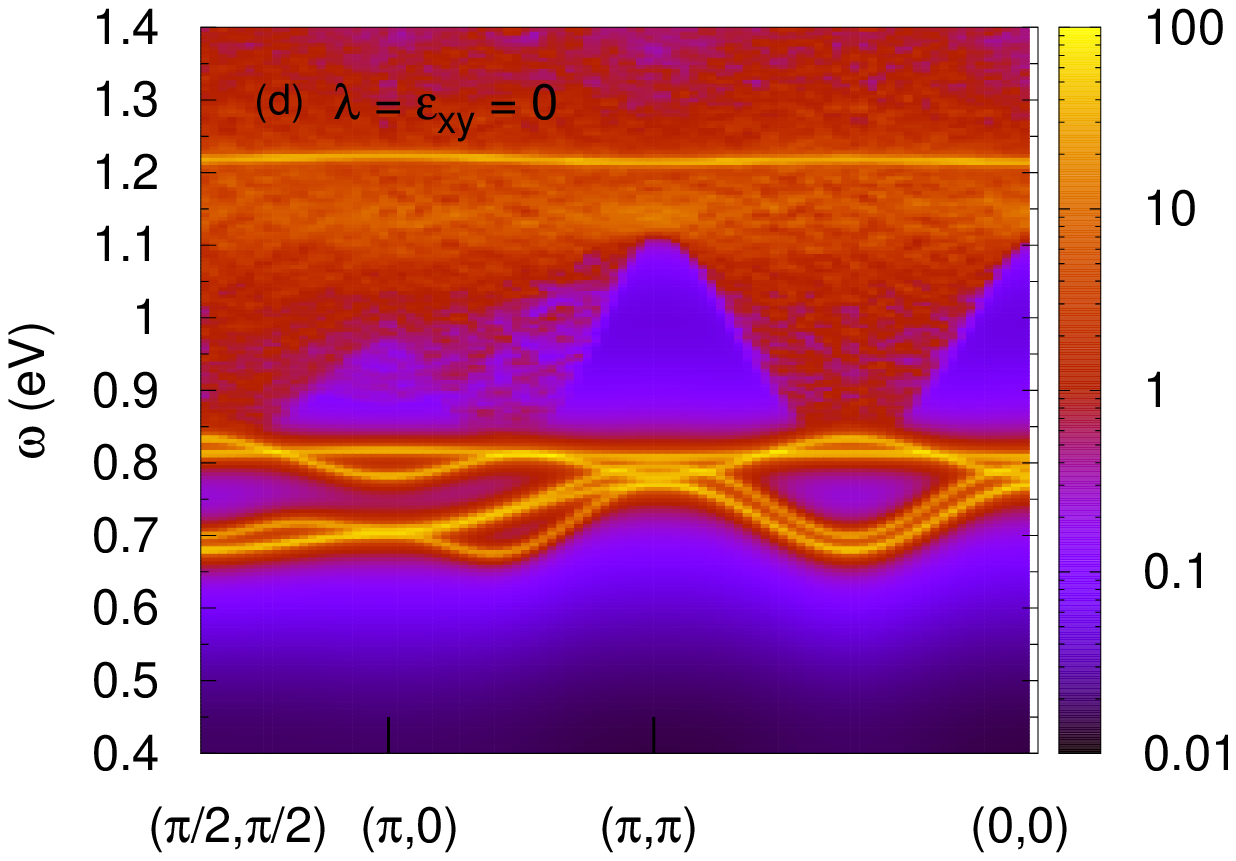,angle=0,width=70mm}
\caption{Gapped magnon modes and dominantly magnetic exciton modes for the $n=3$ case seen in the high-energy part of the spectral function calculated in the self consistent AFM state including octahedral tilting, for different SOC ($\lambda$) values shown in the panels.} 
\label{panels}
\end{figure}

The high energy part of the spectral function is shown in the series of panels in Fig. \ref{panels} for different SOC strengths. The two groups of modes here correspond to: (i) the Hund's coupling induced gapped magnon modes for out-of-phase spin fluctuations (the two dispersive modes starting at energies 0.7 and 0.8 eV from the left edge in panel (a)), and (ii) the spin-orbiton modes (starting at energy below 0.6 eV) which are inter-orbital magnetic excitons corresponding to the lowest-energy particle-hole excitations across the AFM band gap involving $yz/xz$ orbitals (particle) and $xy$ orbital (hole) states (Fig. \ref{band1}(a)). Through the usual resonant scattering mechanism, these modes are pulled down in energy below the continuum by the $U''$ interaction term, and form well defined propagating modes. 

The spin-orbiton mode involving $xy$ orbital shifts to higher energy when $\epsilon_{xy}$ decreases to zero (panel (d)) which lowers the dominantly $xy$ valence band (Fig. \ref{band1}(a)) and thus increases the particle-hole excitation energy. The splitting of the exciton modes in panel (c) is due to the SOC induced splitting of electronic bands as seen in Fig. \ref{band1}(a), which is then reflected in the particle-hole excitation energies. The combination of orbitals for these exciton modes indicates that $L_x$ and $L_y$ components of the orbital angular momentum are involved in these coupled spin-orbital fluctuations. There is an additional spin-orbiton mode involving only $yz,xz$ orbitals (and $L_z$ component) which is formed at higher energy near 0.8 eV (flat band near the left edge iin panel (a)). With increasing SOC, the high-energy modes involving $yz,xz$ orbitals acquire significant spin-orbit exciton character.

\section{$n=5$ --- application to $\rm Sr_2IrO_4$}

The perovskite structured $5d^5$ compound $\rm Sr_2IrO_4$ exhibits an AFM insulating state due to strong SOC induced splitting of the $t_{\rm 2g}$ states, with four electrons in the nominally filled and non-magnetic $J$=3/2 sector and one electron in the nominally half filled and magnetically active $J$=1/2 sector. The SOC induced splitting of $3\lambda/2$ between states of the two total angular momentum sectors, strong spin-orbital entanglement, and band narrowing of states in the $J$=1/2 sector, all of these play a crucial role in the stabilization of the AFM insulator state. Both low-energy magnon excitations and high-energy spin-orbit excitons across the renormalized spin-orbit gap have been intensively studied using RIXS experiments and variety of theoretical approaches.\cite{jkim1_PRL_2012,porras_PRB_2019,mohapatra_PRB_2017,mohapatra_JMMM_2020,mohapatra_JPCM_2021}

\begin{table}[b] \vspace*{-0mm}
\caption{Self consistently determined magnetization and density values, showing small spin canting about the $z$ axis due to octahedral rotation induced DM interaction. Here $t_{m1}=0.2$.} 
\centering % centering table
\begin{tabular}{l c c c c} \\  % creating 5 columns 
\hline\hline % inserting double-line
$\mu$ (s) & $m_\mu^x$ & $m_\mu^y$ & $m_\mu^z$ & $n_\mu$ \\ [0.5ex]
\hline % inserts single-line
$yz$ (A) & 0.186 & $-$0.052 & 0 & 1.653 \\[1ex]
$xz$ (A) & $-$0.185 & 0.049 & 0 & 1.654 \\[1ex]
$xy$ (A) & $-$0.172 & $-$0.047 & 0 & 1.693 \\[1ex] \hline
\end{tabular} \hspace{10mm}
\begin{tabular}{l c c c c} \\  % creating 5 columns 
\hline\hline % inserting double-line
$\mu$ (s) & $m_\mu^x$ & $m_\mu^y$ & $m_\mu^z$ & $n_\mu$ \\ [0.5ex]
\hline % inserts single-line
$yz$ (B) & $-$0.186 & $-$0.052 & 0 & 1.653 \\[1ex]
$xz$ (B) & 0.185 & 0.049 & 0 & 1.654 \\[1ex]
$xy$ (B) & 0.172 & $-$0.047 & 0 & 1.693 \\[1ex] \hline 
% inserts single-line
\end{tabular}
\label{table3}
\end{table}

In this case, we have taken realistic parameter values $U=3$, $J_{\rm H}=U/7$, bare SOC value $\lambda=1.35$, and $\epsilon_{xy}=-0.5$ for simplicity, along with hopping terms: ($t_1$, $t_2$, $t_3$, $t_4$, $t_5$, $t_{m1}$)=(-1.0, 0.5, 0.25, -1.0, 0.0, 0.2), all in units of the realistic hopping energy scale $|t_1|$=290 meV. The self consistently determined results for various physical quantities are given in Table III for magnetic order in the $x$ direction. All ordering directions within the $x-y$ plane are nearly equivalent. Besides the dominant Hund's coupling induced easy-plane anisotropy,\cite{mohapatra_JPCM_2021} there is an extremely weak easy-axis anisotropy which will be discussed at the end of this section. The octahedral rotation induces small in-plane canting of spins but the canting axis is free to orient in any direction. The strong Coulomb interaction induced SOC renormalization by nearly 2/3 (Table IV) agrees with the pseudo-orbital based approach.\cite{mohapatra_JMMM_2020} The strong orbital moments and their correlation with the magnetic order direction (Table IV) reflect the strong SOC induced spin-orbital entanglement. 

\begin{table}[t]
\caption{Self consistently determined renormalized SOC values $\lambda_\alpha = \lambda+\lambda_\alpha^{\rm int}$ and the orbital magnetic moments $\langle L_\alpha \rangle$ for $\alpha=x,y,z$ on the two sublattices. Bare SOC value $\lambda$=1.35.} 
\centering % centering table
\begin{tabular}{l r r r r r r} \\  % creating 6 columns 
\hline\hline % inserting double-line
$s$ & $\lambda_x$ & $\lambda_y$ & $\lambda_z$ & $\langle L_x \rangle$ & $\langle L_y \rangle$ & $\langle L_z \rangle$ \\ [0.5ex]
\hline % inserts single-line
A & 1.882 & 1.882 & 1.871 & 0.367 & 0.091 & 0 \\[1ex]
B & 1.882 & 1.882 & 1.871 & $-$0.367 & 0.091 & 0 \\[1ex]
\hline % inserts single-line
\end{tabular}
\label{table4}
\end{table}

\vspace*{0mm}
\begin{figure}[b]
\hspace*{0mm}
\psfig{figure=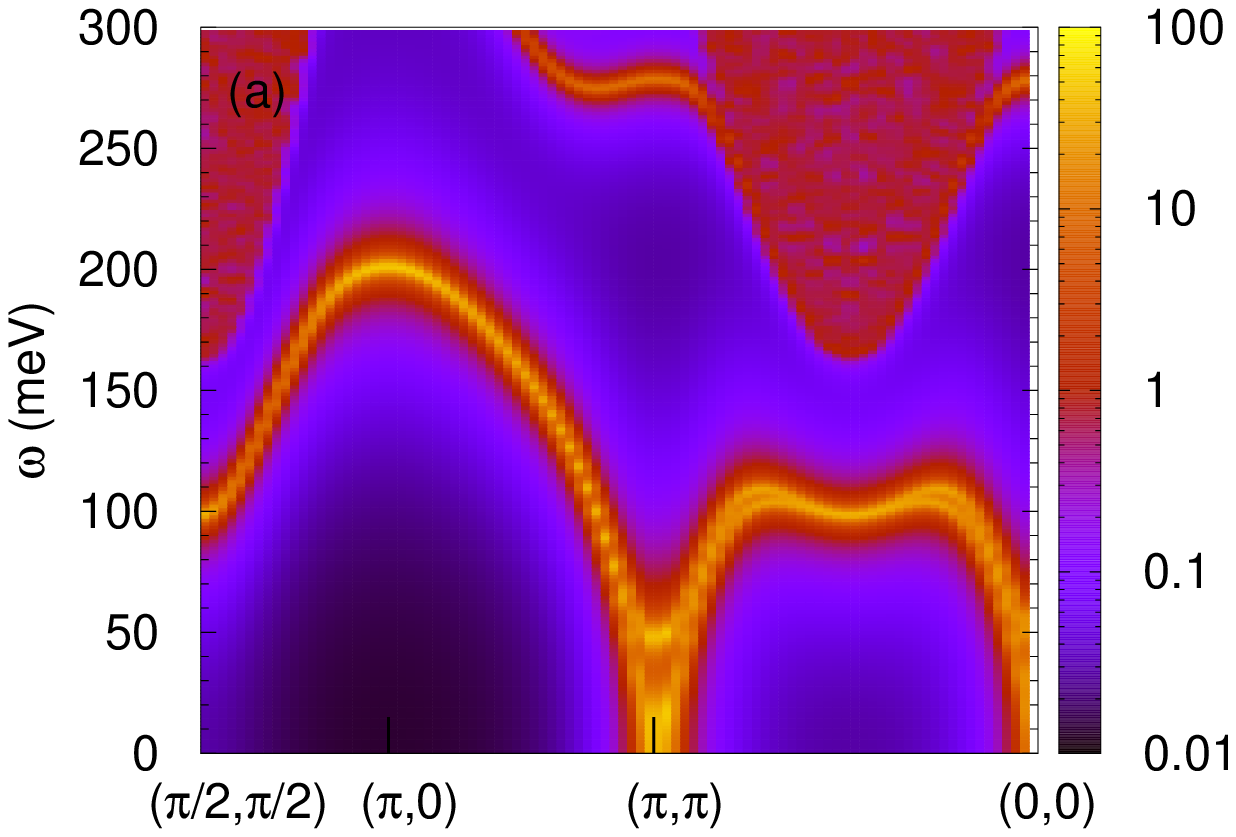,angle=0,width=70mm}
\psfig{figure=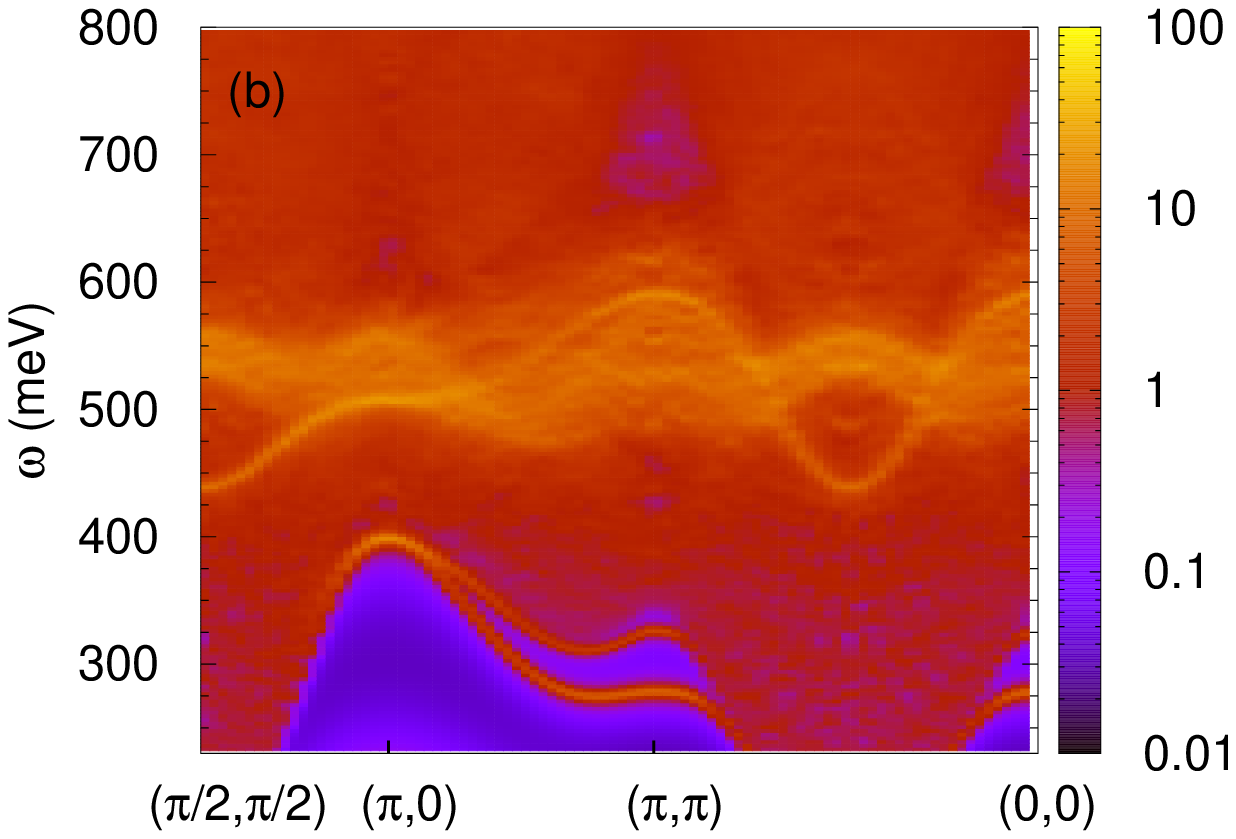,angle=0,width=70mm}
\caption{The spectral function in the self-consistent state for the $n=5$ case with planar AFM order including octahedral rotation, showing the (a) gapless and gapped modes corresponding to in-plane and out-of-plane fluctuations and (b) the spin-orbit exciton modes near 500 meV and 300 meV in the high-energy part.} 
\label{spfn5}
\end{figure}

The low energy part of the spectral function (Fig. \ref{spfn5}(a)) clearly shows the gapless and gapped modes corresponding to in-plane and out-of-plane fluctuations, consistent with the easy-plane anisotropy. The magnon gap $\approx 45$ meV is close to the result obtained using the pseudo-orbital based approach,\cite{mohapatra_JPCM_2021} and in agreement with recent experiments.\cite{porras_PRB_2019} It should be noted that along with the full generalized spin sector, the orbital off-diagonal charge sector ($\psi_\mu^\dagger {\bf 1} \psi_\nu$) related to the orbital moment operators $L_{x,y,z}$ was included in the above calculations which allows for the accompanying transverse fluctuations of orbital moments. Indeed, the exactly gapless Goldstone mode seen in Fig. \ref{spfn5}(a) is obtained only if the $\psi_\mu^\dagger {\bf 1} \psi_\nu$ sector is included, indicating the coupled spin-orbital nature of the Goldstone mode, as illustrated in Appendix C showing the detailed spin-orbital composition.

%, as illustrated in Fig. \ref{panel_n5}
%\begin{figure}[b]
%\hspace*{0mm}
%\psfig{figure=panel_n5.eps,angle=0,width=70mm}
%\caption{Due to the spin-orbital entanglement, magnetic ordering in different directions (a) is the } 
%\label{panel_n5}
%\end{figure}

\begin{figure}
\hspace*{0mm}
\psfig{figure=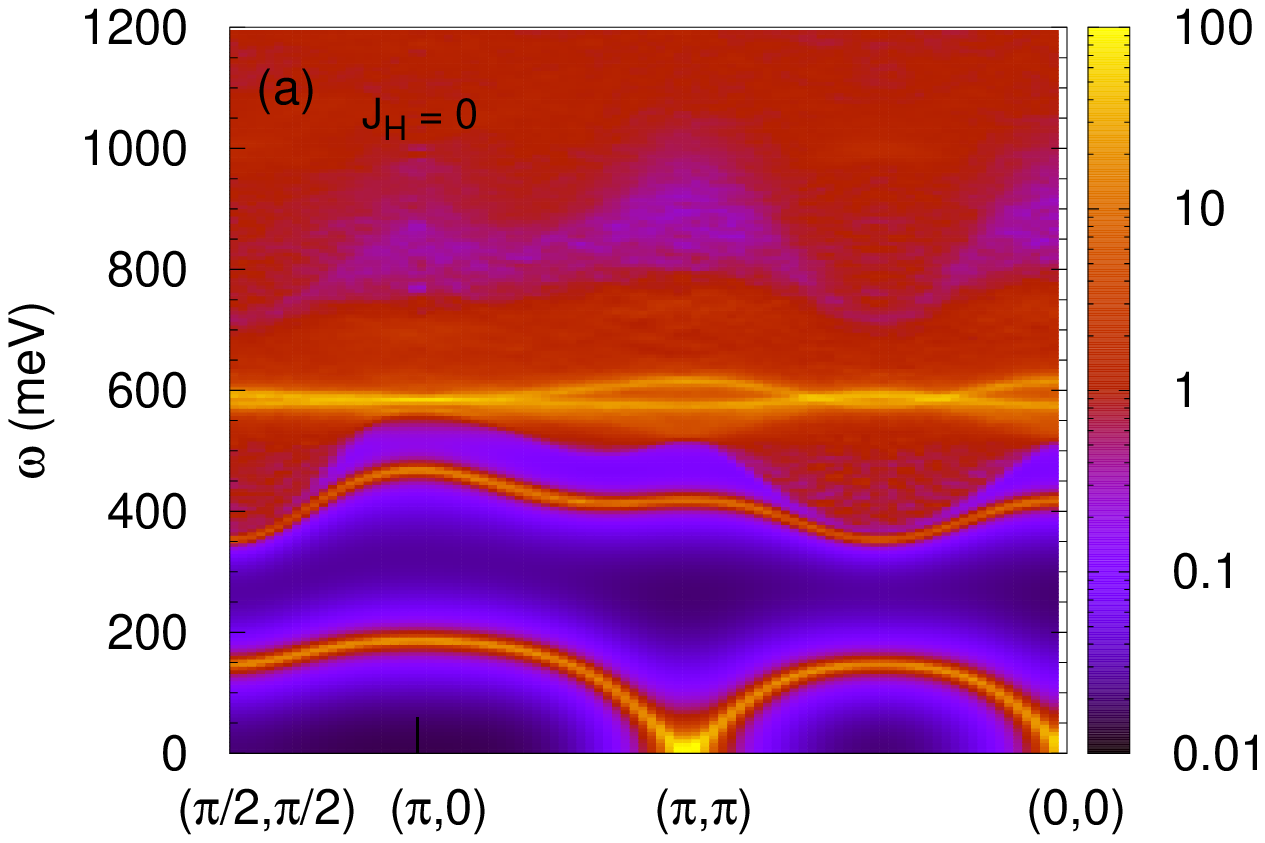,angle=0,width=53mm}
\psfig{figure=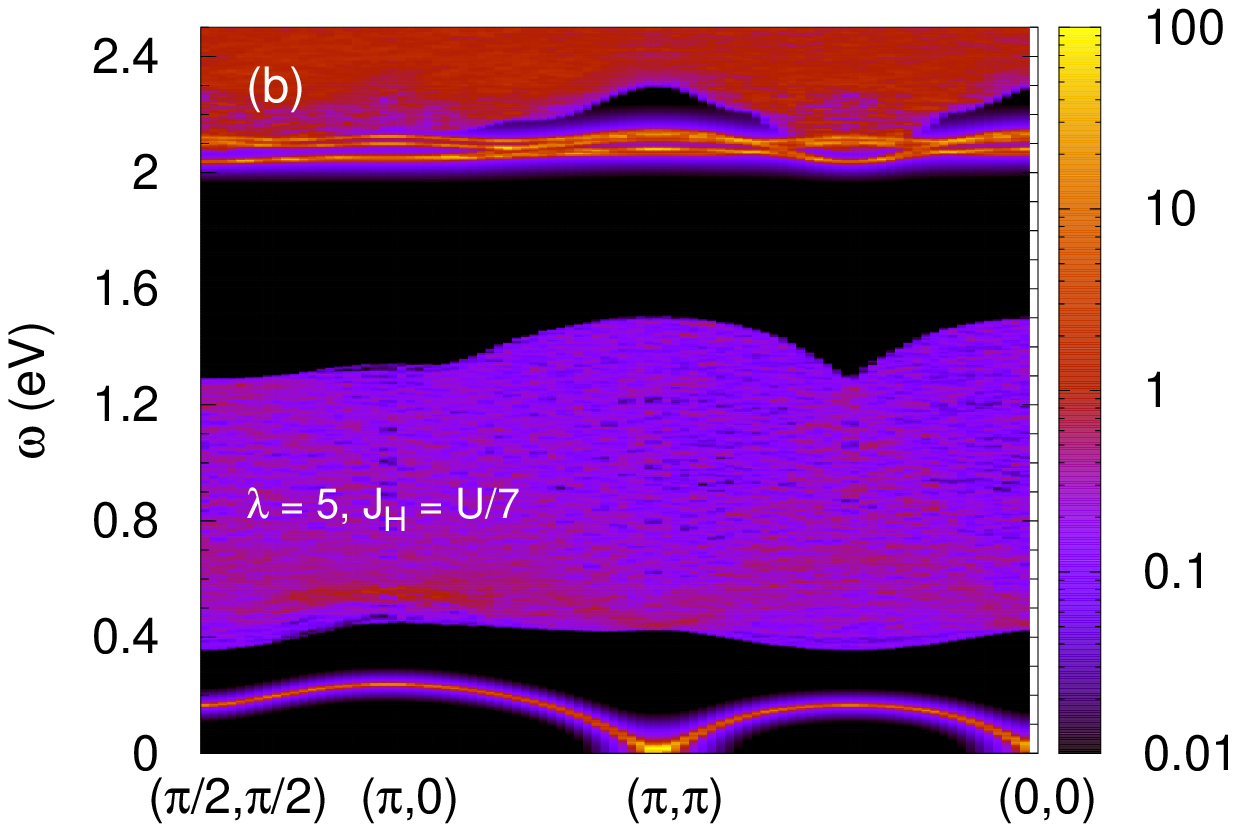,angle=0,width=53mm}
\psfig{figure=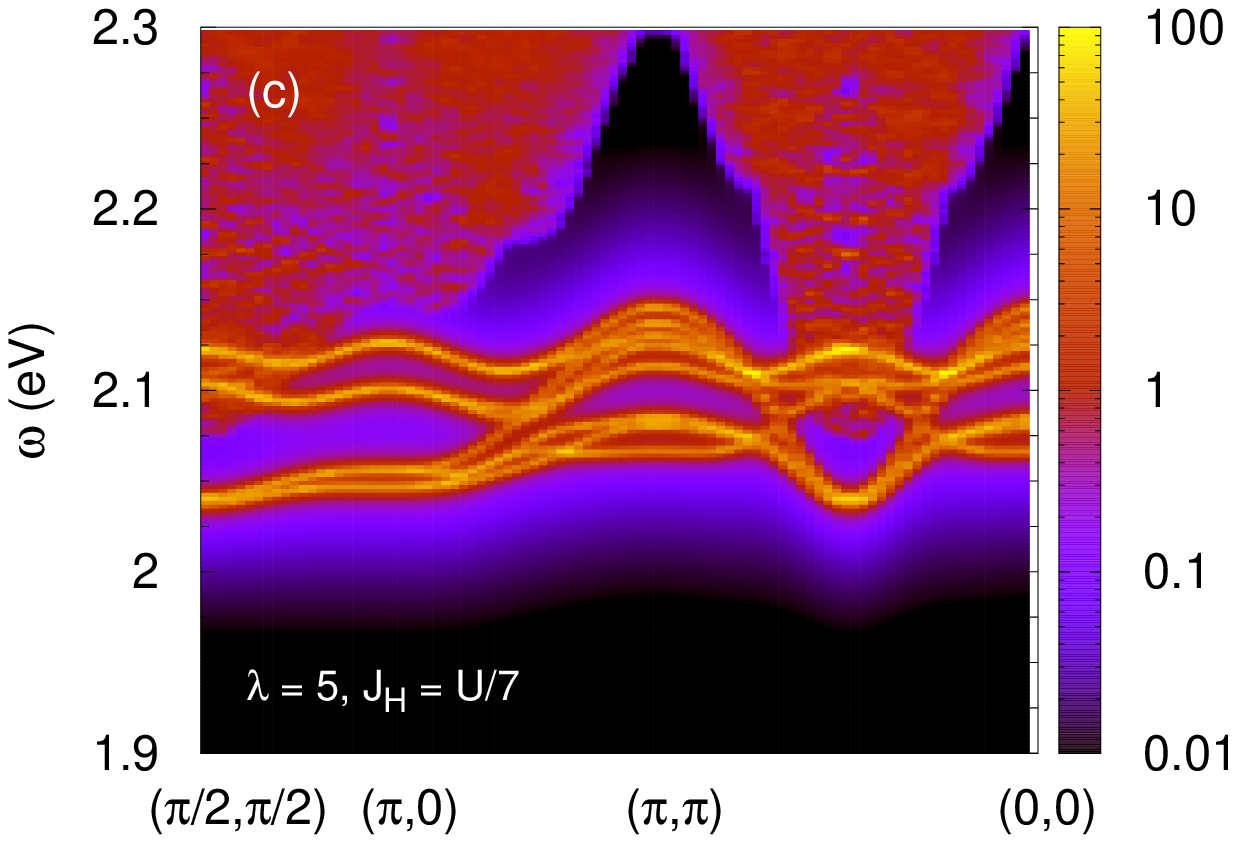,angle=0,width=53mm}
\caption{The spin-orbit exciton modes ($n=5$) for special cases showing (a) no splitting in the weak branch ($\sim$ 400 meV) for Hund's coupling $J_{\rm H}=0$, (b) disappearance of the weak branch for large SOC value $\lambda=5$, and (c) expanded view of the multiple exciton modes ($\sim$ 2 eV) in case (b). Here octahedral rotation is included in all three cases.} 
\label{fig7}
\end{figure}

Fig. \ref{spfn5}(b) shows the spin-orbit exciton modes ($\sim$ 500 meV) involving particle-hole excitations between the $J$=1/2 and 3/2 sectors, which matches closely with results obtained using the pseudo-orbital based approach.\cite{mohapatra_JMMM_2020} As discussed in the previous ($n=3$) case, collective modes arise from particle-hole excitations which are converted to well defined propagating modes split off from the continuum by the Coulomb interaction induced resonant scattering mechanism. The significantly weaker modes ($\sim$ 300 meV) just below the particle-hole continuum for the nominally $J=1/2$ sector are also spin-orbit exciton modes. The splitting seen beyond $(\pi,0)$ vanishes for $J_{\rm H}=0$, as seen in Fig. \ref{fig7}(a). The weak intensity corresponds to the small $J=3/2$ character (mainly $m_J=\pm 3/2$) in the nominally $J=1/2$ bands due to strong mixing between the two sectors induced by the band (hopping) terms. For large SOC strength $\lambda$, the weak exciton modes disappear (Fig. \ref{fig7}(b)), confirming the above picture. Thus, the (low) intensity of the weak exciton modes provides a direct measure of the mixing between the $J=1/2$ and 3/2 sectors. The fine splitting of exciton bands in Fig. \ref{fig7}(c) corresponds to four possible $m_J$ values ($\pm 3/2,\pm 1/2$) for the hole in the $J=3/2$ sector and the exciton hopping terms connecting the two sublattices.  

%corresponds to the two $m_J=\pm 3/2$ cases.

\begin{figure}
\hspace*{0mm} \vspace*{-20mm} 
\psfig{figure=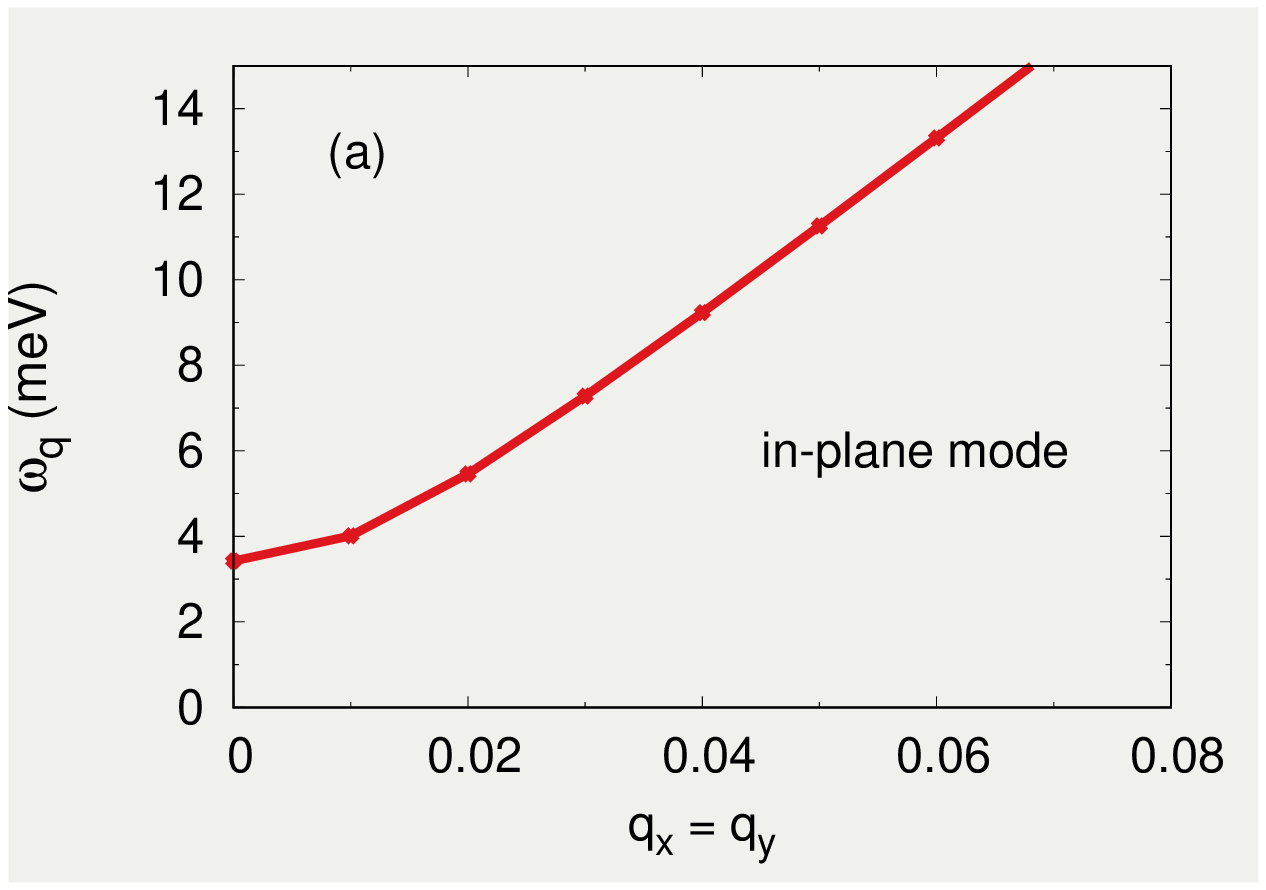,angle=0,width=60mm} \hspace{20mm} 
\psfig{figure=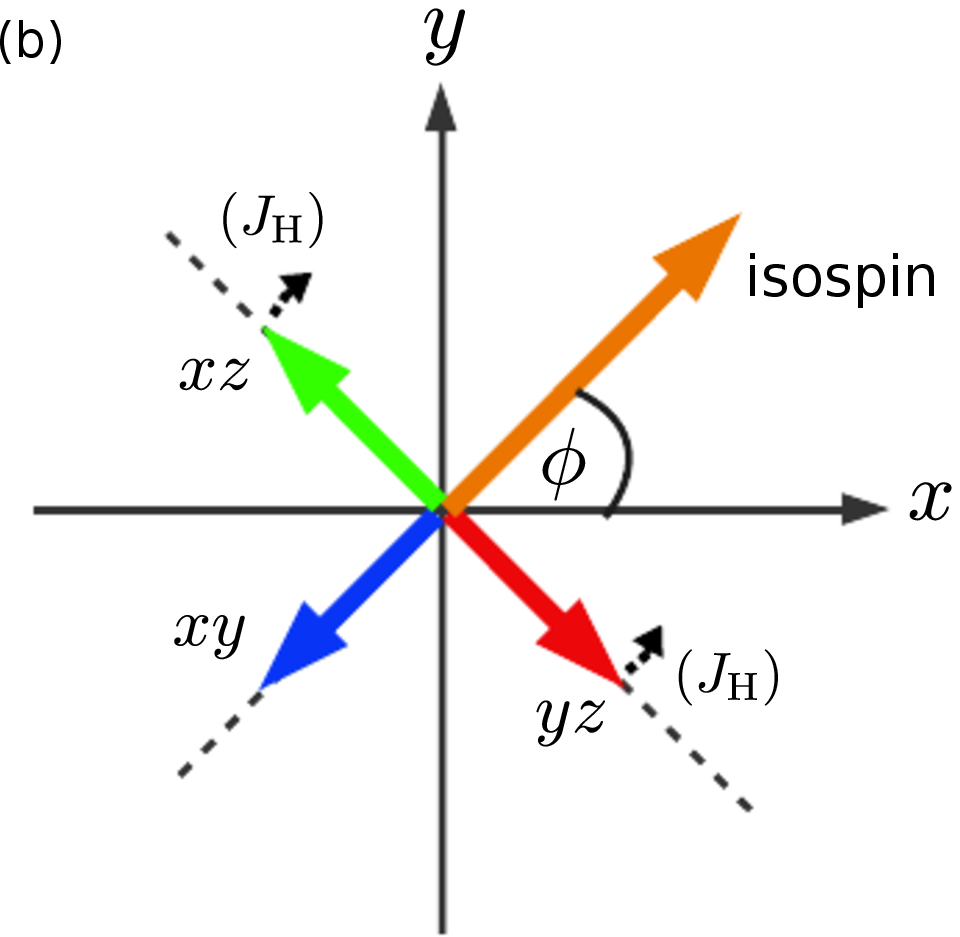,angle=0,width=50mm} \vspace*{20mm}
\caption{(a) Magnon energies for isospin order along $\hat{x}+\hat{y}$ direction showing the small magnon gap ($\approx$ 3 meV) for the in-plane fluctuation mode. (b) The isospin and $yz,xz,xy$ moment orientations for the ideal spin-orbital entangled state, which is extremely weakly perturbed by finite $J_{\rm H}$ resulting in slight twisting of the $yz,xz$ moments as indicated, leading to the easy-axis anisotropy with $C_4$ symmetry. The isospin easy axes are along $\phi=n\pi/4$ where $n=1,3,5,7$.} 
\label{easy_axis}
\end{figure}

We now discuss the extremely weak easy-axis anisotropy which leads to preferred isospin ($J=1/2$) orientation along the diagonal directions $\pm(\hat{x}\pm\hat{y})$ within the easy plane. Fig. \ref{easy_axis}(a) shows the small magnon gap ($\approx$ 3 meV) for the in-plane magnon mode induced by the Hund's coupling $J_{\rm H}$ due to the extremely weak spin twisting as shown in Fig. \ref{easy_axis}(b) which results in an easy-axis anisotropy with $C_4$ symmetry. Here the parameter set is same as earlier including the octahedral rotation which only weakly enhances the magnon gap. The above weak perturbative effect of $J_{\rm H}$ on the strongly spin-orbital entangled state corresponds to the opposite end of the competition between SOC and $J_{\rm H}$ as compared to the $n=3$ case discussed in Sec. V.

%===== n = 4 ======

\section{$n=4$ --- application to $\rm Ca_2RuO_4$}
For moderate tetragonal distortion ($\epsilon_{xy}\approx -1$), the $xy$ orbital in the $4d^4$ compound $\rm Ca_2 Ru O_4$ is nominally doubly occupied and magnetically inactive, while the nominally half-filled and magnetically active $yz,xz$ orbitals yield an effectively two-orbital magnetic system. Hund's coupling between the two $S=1/2$ spins results in low-lying (in-phase) and appreciably gapped (out-of-phase) spin fluctuation modes. The in-phase modes of the $yz,xz$ orbital $S=1/2$ spins correspond to an effective $S=1$ spin system. However, the rich interplay between SOC, Coulomb interaction, octahedral rotations, and tetragonal distortion results in complex magnetic behaviour which crucially involves the $xy$ orbital and is therefore beyond the above simplistic picture. 

Treating all the different physical elements on the same footing within the unified framework of the generalized self-consistent approach explicitly shows the variety of physical effects arising from the rich interplay in $\rm Ca_2RuO_4$. These include: SOC induced easy-plane and easy-axis anisotropies similar to the $n$=3 case, octahedral tilting induced reduction of easy-axis anisotropy from $C_4$ to $C_2$ symmetry, spin-orbital coupling induced orbital magnetic moments, Coulomb interaction induced strongly anisotropic SOC renormalization, decreasing tetragonal distortion induced magnetic reorientation transition from planar AFM order to FM ($z$) order, and orbital moment interaction induced orbital gap.\cite{mohapatra_JPCM_2020} Stable FM and AFM metallic states were also obtained near the magnetic phase boundary separating the two magnetic orders. The self-consistent determination of magnetic order has also explicitly shown the coupled nature of spin and orbital fluctuations, as reflected in the ferro and antiferro orbital fluctuations associated with in-phase and out-of-phase spin twisting modes, highlighting the strong deviation from conventional Heisenberg behaviour in effective spin models, as discussed recently to account for the magnetic excitation measurements in INS experiments on $\rm Ca_2 Ru O_4$.\cite{jain_NATPHY_2017} 

In the following, we will take the same parameter set as considered in the self-consistent study,\cite{mohapatra_JPCM_2020} along with $U=8$ and $J_{\rm H}=U/5$ in the energy scale unit (150 meV), so that $U=1.2$ eV, $U^{\prime\prime}=U/2=0.6$ eV, and $J_{\rm H}=0.24$ eV. These are comparable to reported values extracted from RIXS ($J_{\rm H}=0.34$ eV) and ARPES ($J_{\rm H}=0.4$ eV) studies.\cite{gretarsson_PRB_2019,sutter_NATCOM_2017} The hopping parameter values considered are as given in Sec. II, and the bare SOC value $\lambda=1$.

%Since the orbital off-diagonal condensates contribute on the same footing as the normal condensates, the coupled spin-charge-orbital fluctuations must be investigated within a unified formalism involving the generalized spin and charge operators, as discussed below.

%The magnetization and density values for the three orbitals are presented in Table I, all off-diagonal spin and charge condensates in Table II, and the renormalized SOC values and orbital magnetic moments in Table III. Here $U=8$, $\epsilon_{xy}=-0.8$, the bare SOC strength $\lambda^{\rm bare}=1$, and the staggered octahedral rotation ($t_{m1}=0.2$) and tilting ($t_{m2}=t_{m3}=0.15$) have been included. 

%Similarly, the magnetic anisotropy effect of SOC induced $xy$-orbital density change on spin twisting from $z$ direction to $x-y$ plane, which couples to the tetragonal distortion induced orbital energy offset $\epsilon_{xy}$, can be investigated on the magnon gap in terms of spin-charge coupling. 

\begin{figure}
\hspace*{0mm}
\psfig{figure=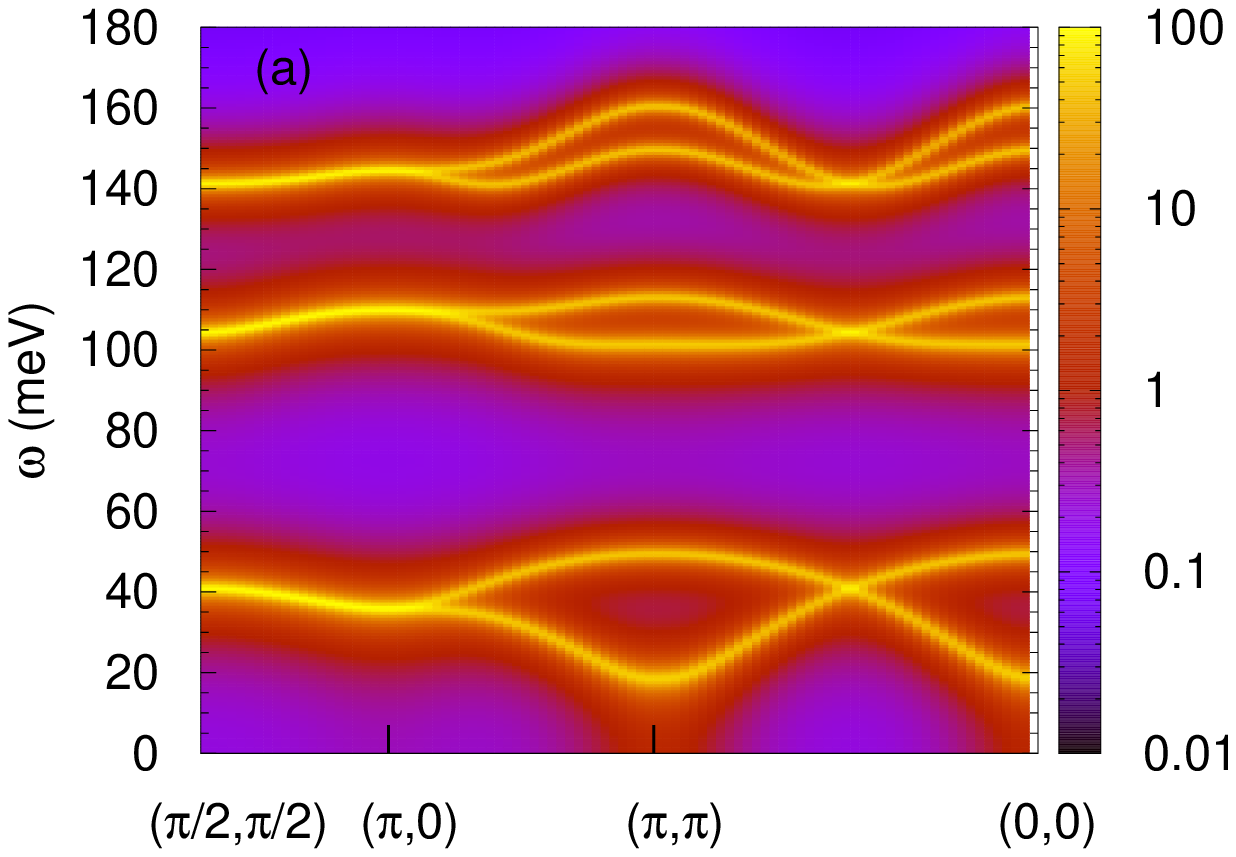,angle=0,width=80mm}
\psfig{figure=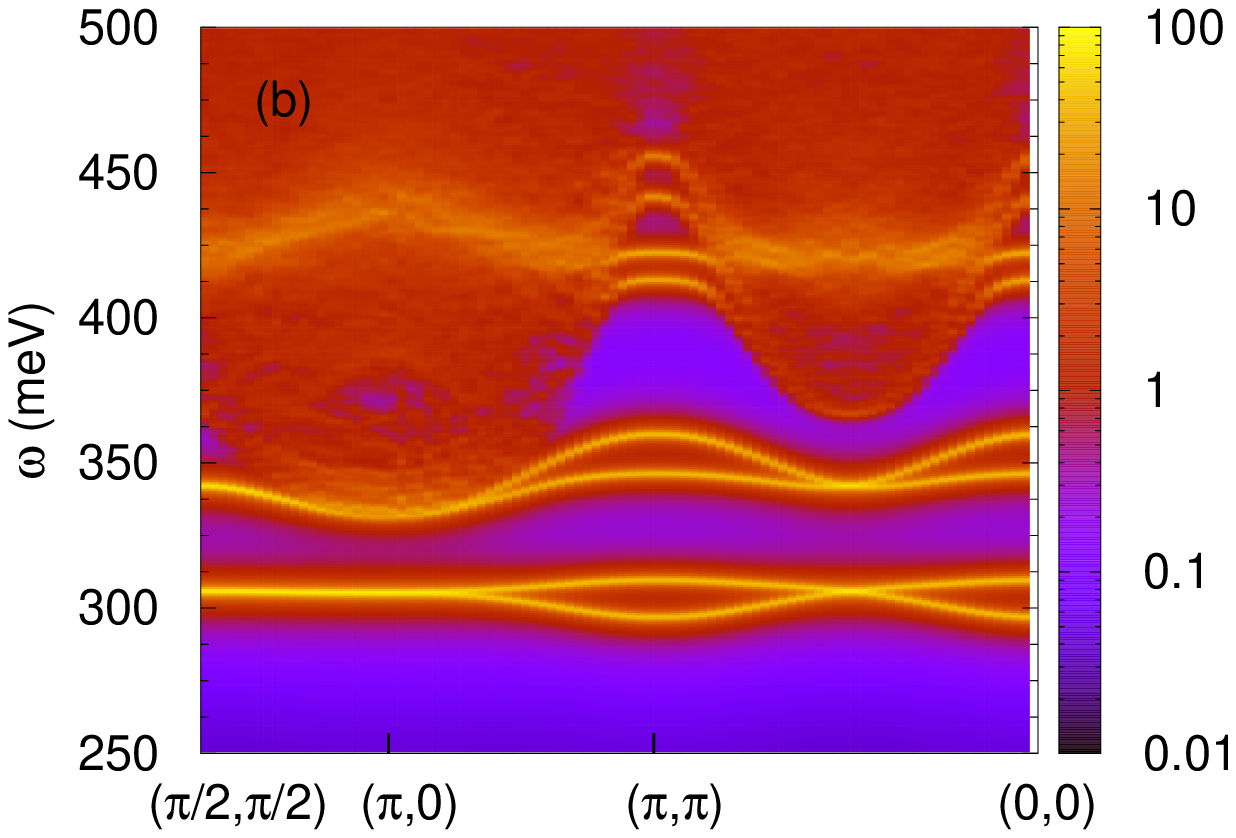,angle=0,width=80mm}
\caption{The generalized fluctuation spectral function for the $n=4$ case, showing coupled spin-orbital excitations including low-energy magnon modes (below $\sim$ 60 meV), intermediate-energy orbiton (100 and 140 meV) and spin-orbiton (300 and 350 meV) modes, and high-energy spin-orbit exciton (425 meV) modes.} 
\label{spfn_n4}
\end{figure}

%Please find the attached new figures named according to the figure numbers in the manuscript.
%Fig8a and 8b (ruthenate SOC=1 and exy = -0.8 case) are plotted with reduced 150 meV energy scale.
%We may consider including the higher energy modes (> 500 meV) in the manuscript as these modes with reduced energy 
%scale are in agreement with RIXS data. Therefore I am attaching this as Fig8c.
%I will soon send the remaining fig7 panel and easy-axis magnon gap plot (with one para description).

Fig. \ref{spfn_n4} shows the calculated generalized fluctuation spectral function. Several well defined propagating modes are seen here including: (i) the low-energy (below $\sim$ 60 meV) dominantly spin (magnon) excitations involving the magnetically active $yz,xz$ orbitals and corresponding to in-plane and out-of-plane fluctuations which are gapped due to the magnetic anisotropies, (ii) the intermediate-energy (100 and 140 meV) dominantly orbital excitations (orbitons) involving particle-hole excitations between $xy$ (hole) and $yz,xz$ (particle) states, (iii) the intermediate-energy (300 and 350 meV) dominantly spin-orbital excitations (spin-orbitons) involving $xy$ (hole) and $yz,xz$ (particle) states, and (iv) the high-energy (425 meV) dominantly spin-orbital excitations (spin-orbit excitons) involving particle-hole excitations between $yz,xz$ (hole) and $yz,xz$ (particle) states of nominally different $J$ sectors. The SOC-induced spin-orbital entangled $J$ states are strongly renormalized by the tetragonal splitting and the electronic correlation induced staggered field. 

The spin-orbital characterization of the various collective excitations mentioned above is inferred from the basis-resolved contributions to the total spectral functions which explicitly show the relative spin-orbital composition of the various excitations (Appendix C). The presence of sharply defined collective excitations for the magnon, orbiton, and spin-orbiton modes which are clearly separated from the particle-hole continuum highlights the rich spin-orbital physics in the $n=4$ case corresponding to the $\rm Ca_2RuO_4$ compound. Many of our calculated magnon spectra features such as the magnon gaps for in-plane and out-of-plane modes, weak dispersive nature along the magnetic zone boundary, as well as the overall magnon energy scale are in excellent agreement with the INS study.\cite{kunkemoller_PRL_2015,jain_NATPHY_2017} The orbiton mode energy scale is also qualitatively comparable to the composite excitation peaks obtained around 80 meV in Raman and RIXS studies.\cite{souliou_PRL_2017,fatuzzo_PRB_2015,das_PRX_2018,gretarsson_PRB_2019} The calculated spin-orbiton and spin-orbit exciton energies are also in agreement with the excitation peaks obtained around 300-350 meV energy range and 400 meV in RIXS studies. We also obtained excitations in the high-energy range 750-800 meV and 900 meV (not shown), which are comparable to the peaks obtained around 750 meV and 1000 meV in RIXS studies.

%Our calculated spectral function also shows high-energy excitations near 800 and 900 meV, as also observed in RIXS studies.

\newpage
\section{Conclusions}
Following up on the generalized self-consistent approach including orbital off-diagonal spin and charge condensates, investigation of the generalized fluctuation propagator reveals the composite spin-orbital character of the different types of collective excitations in strongly spin-orbit coupled systems. A realistic representation of magnetic anisotropy effects due to the interplay of SOC, Coulomb interaction, and structural distortion terms was included in the three-orbital model, while maintaining uniformity of lattice structure in order to focus on the coupled spin-orbital excitations. Our unified investigation of the three electron filling cases $n=3,4,5$ corresponding to the three compounds $\rm NaOsO_3$, $\rm Ca_2RuO_4$, $\rm Sr_2IrO_4$ provides deep insight into how the spin-orbital physics in the magnetic ground state is reflected in the collective excitations. The calculated spectral functions show well defined propagating modes corresponding to dominantly spin (magnon), orbital (orbiton), and spin-orbital (spin-orbiton) excitations, along with the spin-orbit exciton modes involving spin-orbital excitations between states of different $J$ sectors induced by the spin-orbit coupling. 

\appendix
\section{Orbital off-diagonal condensates in the HF approximation}
The additional contributions in the HF approximation arising from the orbital off-diagonal spin and charge condensates are given below. For the density, Hund's coupling, and pair hopping interaction terms in Eq. \ref{h_int}, we obtain (for site $i$):
\begin{eqnarray}
U'' \sum_{\mu < \nu} n_\mu n_\nu & \rightarrow & 
-\frac{U''}{2} \sum_{\mu < \nu} \left [ n_{\mu\nu} \langle n_{\nu\mu} \rangle + \makebox{\boldmath $\sigma$}_{\mu\nu}.\langle \makebox{\boldmath $\sigma$}_{\nu\mu} \rangle \right ] + {\rm H.c.} \nonumber \\
-2J_{\rm H} \sum_{\mu < \nu} {\bf S}_\mu . {\bf S}_\nu 
& \rightarrow & \frac{J_{\rm H}}{4} \sum_{\mu < \nu} 
\left [3\, n_{\mu\nu} \langle n_{\nu\mu} \rangle - \makebox{\boldmath $\sigma$}_{\mu\nu}.\langle \makebox{\boldmath $\sigma$}_{\nu\mu} \rangle \right ] + {\rm H.c.} \nonumber \\
J_{\rm P} \sum_{\mu \ne \nu} a_{\mu\uparrow}^{\dagger} a_{\mu\downarrow}^{\dagger} a_{\nu\downarrow} a_{\nu\uparrow} & \rightarrow &  
\frac{J_{\rm P}}{2} \sum_{\mu < \nu} \left [n_{\mu\nu} \langle n_{\mu\nu} \rangle - \makebox{\boldmath $\sigma$}_{\mu\nu}.\langle \makebox{\boldmath $\sigma$}_{\mu\nu} \rangle \right ] + {\rm H.c.} 
\end{eqnarray}
in terms of the orbital off-diagonal spin ($\makebox{\boldmath $\sigma$}_{\mu\nu}=\psi_\mu^\dagger \makebox{\boldmath $\sigma$} \psi_\nu$) and charge ($n_{\mu\nu} = \psi_\mu^\dagger {\bf 1} \psi_\nu$) operators. The orbital off-diagonal condensates are finite due to the SOC-induced spin-orbital correlations. These additional terms in the HF theory explicitly preserve the SU(2) spin rotation symmetry of the various Coulomb interaction terms. 

Collecting all the spin and charge terms together, we obtain the orbital off-diagonal (OOD) contributions of the Coulomb interaction terms:
\begin{eqnarray}
[{\cal H}_{\rm int}^{\rm HF}]_{\rm OOD} &=& \sum_{\mu < \nu} \left [ \left (-\frac{U''}{2} + \frac{3J_{\rm H}}{4} \right ) 
n_{\mu\nu} \langle n_{\nu\mu} \rangle + \left (\frac{J_{\rm P}}{2}\right ) n_{\mu\nu} \langle n_{\mu\nu} \rangle \right . \nonumber \\
& & - \left . \left (\frac{U''}{2} + \frac{J_{\rm H}}{4} \right )\makebox{\boldmath $\sigma$}_{\mu\nu}.\langle \makebox{\boldmath $\sigma$}_{\nu\mu} \rangle - \left (\frac{J_{\rm P}}{2} \right )\makebox{\boldmath $\sigma$}_{\mu\nu}.\langle \makebox{\boldmath $\sigma$}_{\mu\nu} \rangle \right ] + {\rm H.c.} 
\end{eqnarray}

\section{Coulomb interaction matrix elements in the orbital-pair basis}
Corresponding to the above HF contributions in the orbital off-diagonal sector, we express the Coulomb interactions in terms of the generalized spin and charge operators (for site $i$):
\begin{eqnarray}
[{\cal H}_{\rm int}]_{\rm OOD} &=& \sum_{\mu < \nu} \left [ \left (-\frac{U''}{2} + \frac{3J_{\rm H}}{4} \right ) 
n_{\mu\nu} n_{\mu\nu}^\dagger - \left (\frac{U''}{2} + \frac{J_{\rm H}}{4} \right )\makebox{\boldmath $\sigma$}_{\mu\nu}. \makebox{\boldmath $\sigma$}_{\mu\nu}^\dagger \right ] \nonumber \\
& + & \sum_{\mu < \nu} \left [ \left (\frac{J_{\rm P}}{4}\right ) n_{\mu\nu} n_{\nu\mu}^\dagger - \left (\frac{J_{\rm P}}{4} \right )\makebox{\boldmath $\sigma$}_{\mu\nu}. \makebox{\boldmath $\sigma$}_{\nu\mu}^\dagger + {\rm H.c.} \right ] 
\label{int_anom}
\end{eqnarray}
where $n_{\mu\nu}^\dagger = n_{\nu\mu}$ and $\makebox{\boldmath $\sigma$}_{\mu\nu}^\dagger = \makebox{\boldmath $\sigma$}_{\nu\mu}$. The above form shows that only the pair-hopping interaction terms $(J_{\rm P})$ are off-diagonal in the orbital-pair $(\mu\nu)$ basis. We will use the above Coulomb interaction terms in the orbital off-diagonal sector in the RPA series in order to ensure consistency with the self-consistent determination of magnetic order including the orbital off-diagonal condensates.  

The Coulomb interaction terms in the orbital diagonal sector can be cast in a similar form:
\begin{equation}
[{\cal H}_{\rm int}]_{\rm OD} = \sum_\mu \left [\left ( -\frac{U}{4} \right ) \makebox{\boldmath $\sigma$}_\mu . \makebox{\boldmath $\sigma$}_\mu +  \left (\frac{U}{4}\right ) n_\mu n_\mu \right ] 
+ \sum_{\mu < \nu} \left [ \left (-\frac{2J_{\rm H}}{4} \right ) \makebox{\boldmath $\sigma$}_\mu . \makebox{\boldmath $\sigma$}_\nu + U'' n_\mu n_\nu \right ]
\label{int_normal}
\end{equation}
which include the Hubbard, Hund's coupling, and density interaction terms. 

The form of the $[U]$ matrix used in the RPA series Eq. (\ref{rpa}) is now discussed below. In the composite spin-charge-orbital-sublattice ($\mu\nu\alpha s$) basis, the $[U]$ matrix is diagonal in spin, charge, and sublattice sectors. There are two possible cases involving the orbital-pair $(\mu\nu)$ basis. In the case $\mu=\nu$, the $[U]$ matrices in the spin ($\alpha=x,y,z$) and charge ($\alpha=c$) sectors are obtained as:
\begin{equation}
[U]_{\mu \mu \alpha=x,y,z}^{\mu'\mu'\alpha'=\alpha} = 
\left [ \begin{array}{ccc} 
U & J_{\rm H} & J_{\rm H} \\ 
J_{\rm H} & U & J_{\rm H} \\ 
J_{\rm H} & J_{\rm H} & U
\end{array} \right ] \hspace{0.5in} 
[U]_{\mu \mu \alpha=c}^{\mu'\mu'\alpha'=\alpha} = 
\left [ \begin{array}{ccc} 
-U & -2U'' & -2U'' \\ 
-2U'' & -U & -2U'' \\ 
-2U'' & -2U'' & -U
\end{array} \right ] 
\end{equation} 
corresponding to the interaction terms (Eq. \ref{int_normal}) for the normal spin and charge density operators. Similarly, for the six orbital-pair cases $(\mu,\nu)$ corresponding to $\mu \ne \nu$, the $[U]$ matrix elements in the spin ($\alpha=x,y,z$) and charge ($\alpha=c$) sectors are obtained as:
\begin{eqnarray}
& & [U]_{\mu\nu\alpha=x,y,z}^{\mu\nu\alpha} = U'' + J_{\rm H}/2 \hspace{0.5in} 
\; [U]_{\mu\nu\alpha=c}^{\mu\nu\alpha} = U'' - 3J_{\rm H}/2  \nonumber \\
& & [U]_{\mu\nu\alpha=x,y,z}^{\nu\mu\alpha} = J_{\rm P} \hspace{1.08in} 
\; [U]_{\mu\nu\alpha=c}^{\nu\mu\alpha} = -J_{\rm P} 
\end{eqnarray}
corresponding to the interaction terms (Eq. \ref{int_anom}) involving the orbital off-diagonal spin and charge operators. 

\begin{figure}[t]
\vspace*{-10mm}
\hspace*{0mm}
\psfig{figure=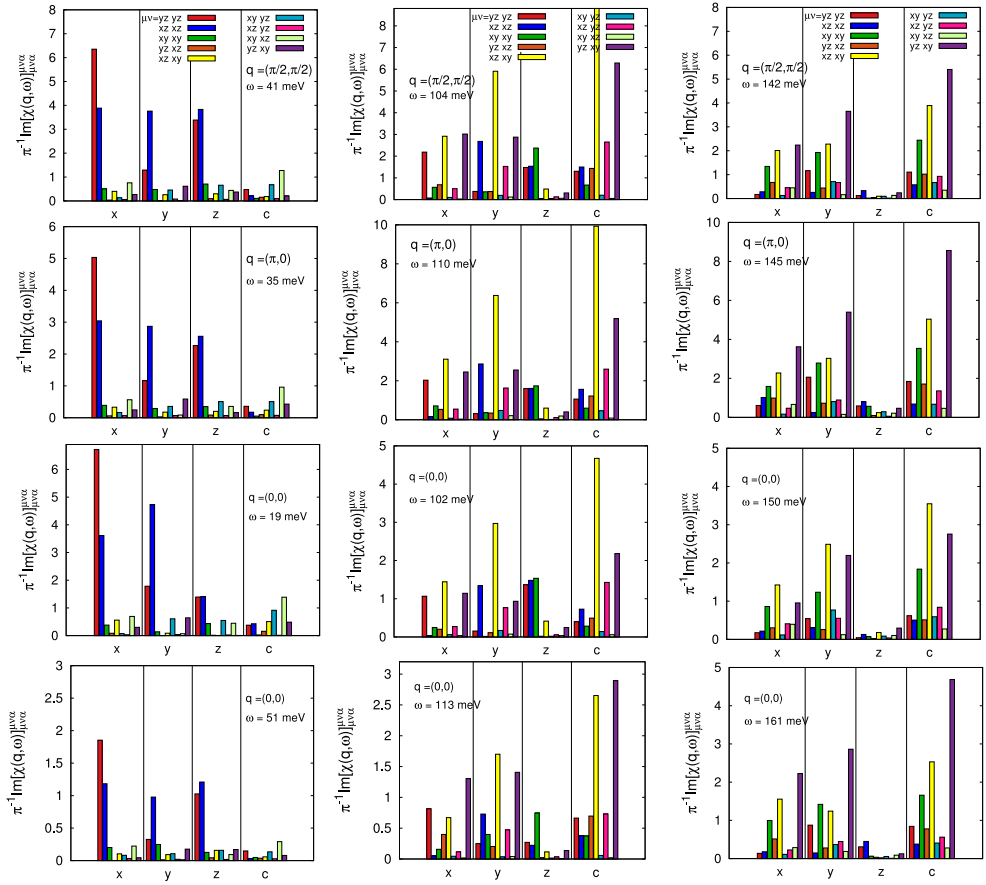,angle=0,width=130mm}
\caption{The basis-resolved contributions to the total spectral function for the low-energy magnon (left panel) and intermediate-energy orbiton (center and right panels) modes, showing dominantly spin ($\mu=\nu$, $\alpha=x,y,z$) and orbital ($\mu\ne\nu$, $\alpha=c$) character of the fluctuation modes, respectively.} 
\label{histo_1}
\end{figure}

\section{Basis-resolved contributions to the total spectral function}
\begin{figure}[t]
\vspace*{10mm}
\hspace*{0mm}
\psfig{figure=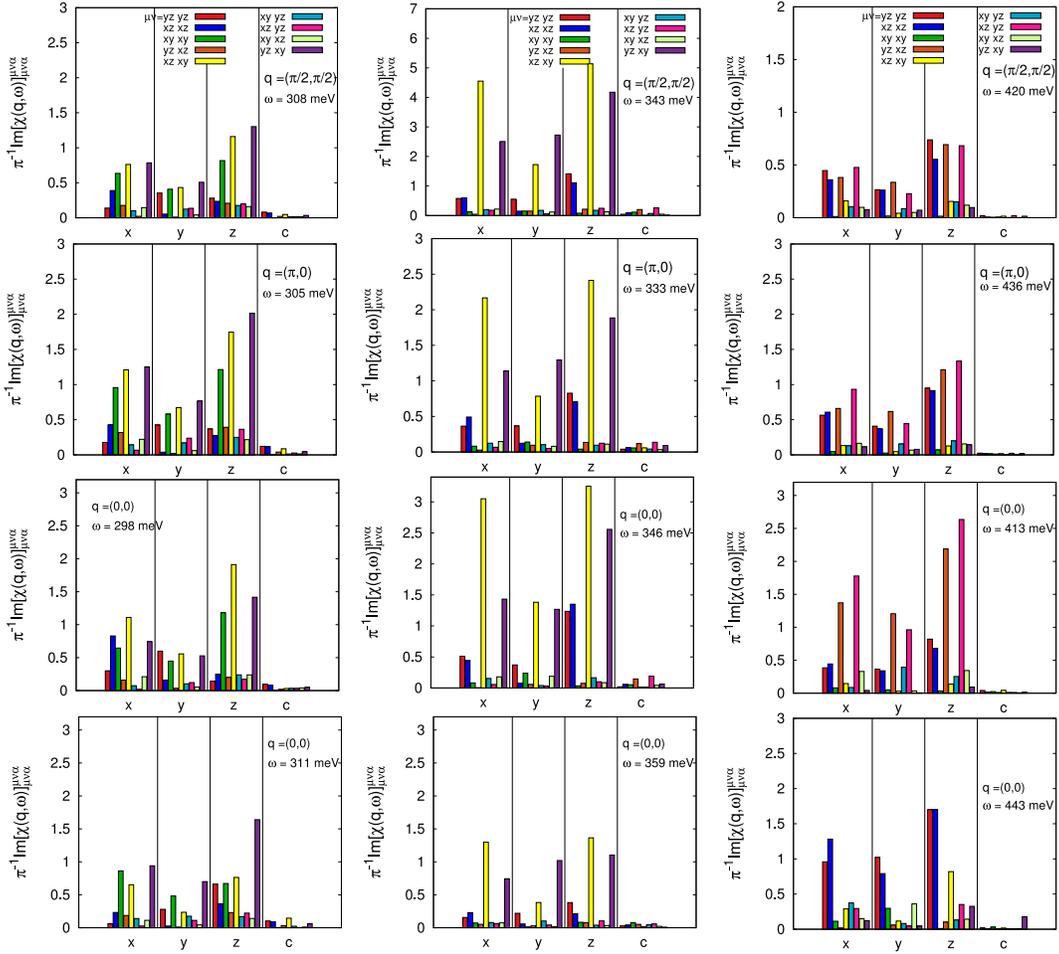,angle=0,width=140mm}
\caption{The basis-resolved contributions to the total spectral function for the intermediate-energy spin-orbiton (left and center panels) and high-energy spin-orbit exciton (right panel) modes, showing dominantly spin-orbital character ($\mu\ne\nu$, $\alpha=x,y,z$) involving $xy$ and $yz,xz$ orbitals (left and center panels) and $yz,xz$ orbitals (right panel).} 
\label{histo_2}
\end{figure}

The detailed spin-orbital character of the collective excitations can be identified from the basis-resolved contributions to the total spectral functions. This is illustrated here for the excitations shown in Fig. \ref{spfn_n4} for the $n=4$ case corresponding to the $\rm Ca_2RuO_4$ compound. Fig. \ref{histo_1} shows dominantly spin excitations involving $yz,xz$ orbitals for the magnon modes (below 60 meV) and dominantly orbital excitations involving $xy$ and $yz,xz$ orbitals for the orbiton modes (100 and 140 meV). Similarly, Fig. \ref{histo_2} shows dominantly spin-orbital excitations involving $xy$ and $yz,xz$ orbitals for the spin-orbiton modes (300 and 350 meV), and dominantly spin-orbital excitations involving $yz,xz$ orbitals for the spin-orbit exciton modes (425 meV). 

\begin{figure}
\vspace*{0mm}
\hspace*{0mm}
\psfig{figure=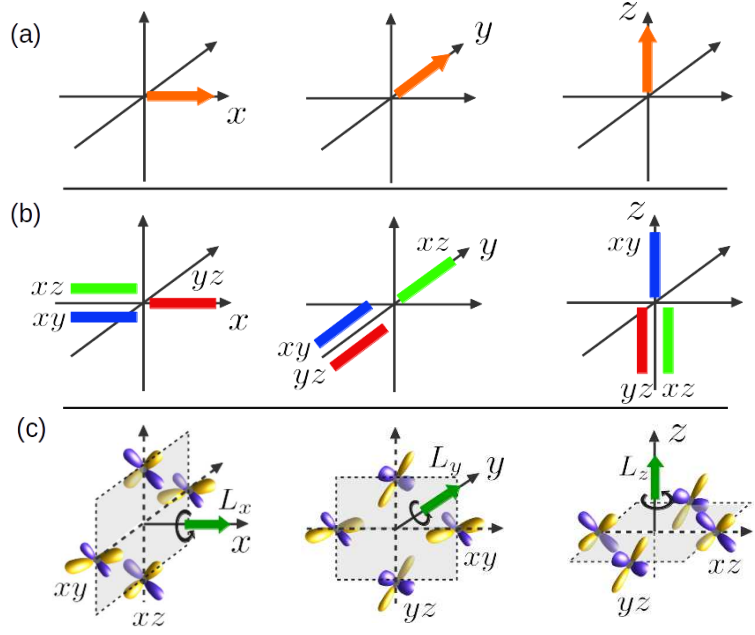,angle=0,width=100mm}
\caption{The extreme spin-orbital-entanglement induced correspondence between (a) magnetic ordering directions, (b) sign of magnetic moments for the three orbitals, and (c) orbital current induced orbital moments for the three orbitals, for the $n=5$ case corresponding to $\rm Sr_2IrO_4$.} 
\label{paneln5}
\end{figure}

\begin{figure}
\vspace*{0mm}
\hspace*{0mm}
\psfig{figure=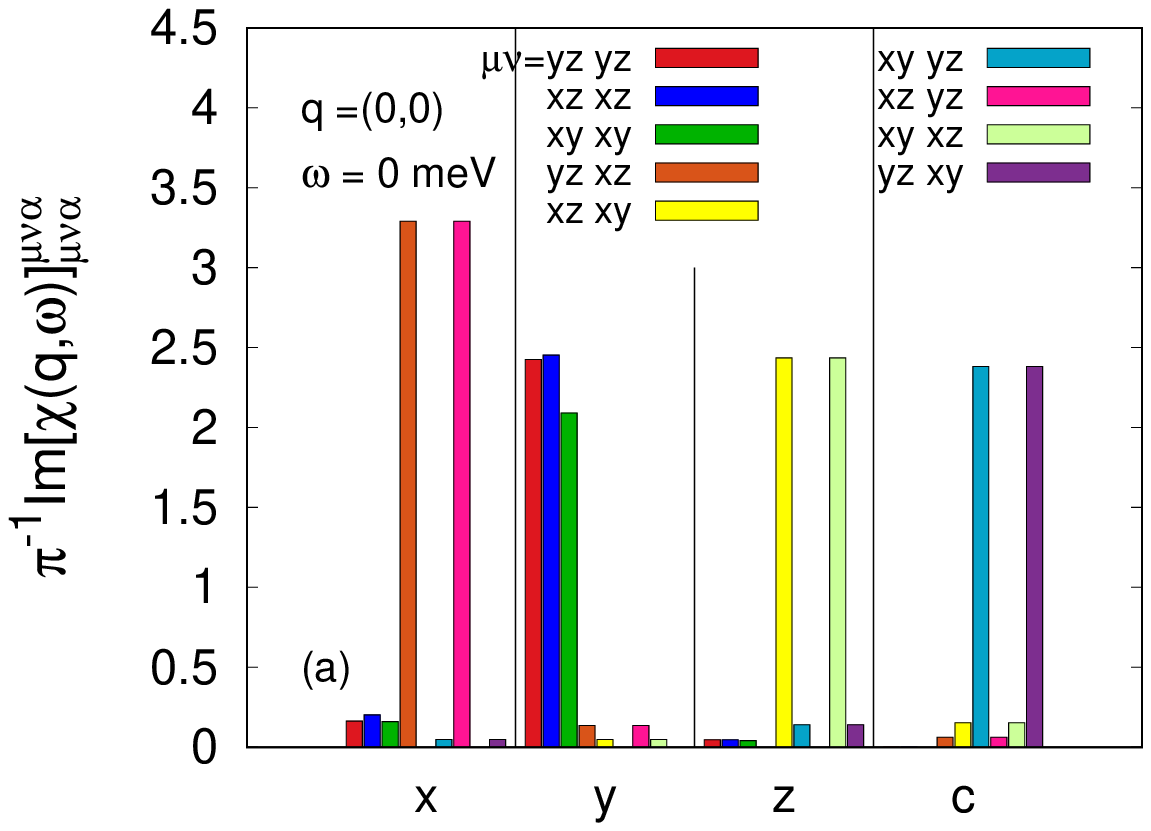,angle=0,width=70mm}
\psfig{figure=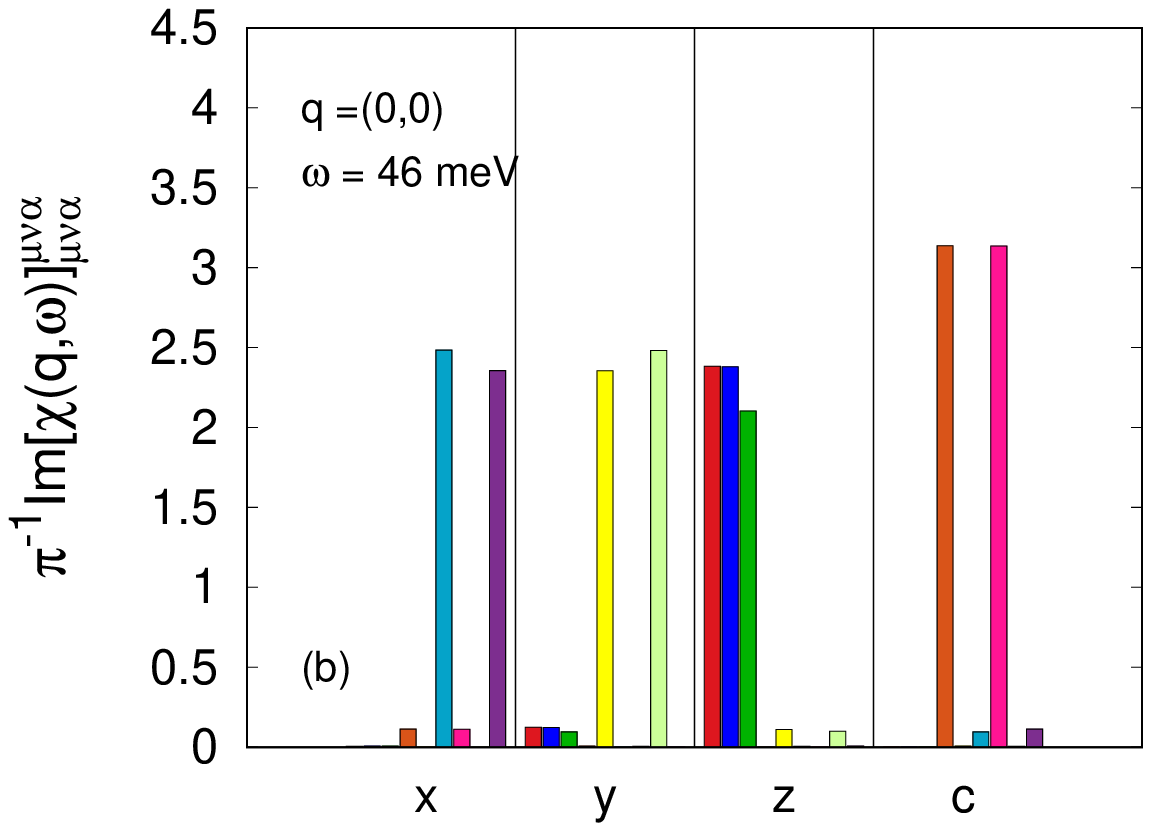,angle=0,width=70mm}
\caption{The basis-resolved contributions to the total spectral function for the (a) gapless in-plane magnon mode and (b) gapped out-of-plane magnon mode for the $n=5$ case corresponding to $\rm Sr_2IrO_4$ with extreme spin-orbital entanglement.} 
\label{histo_3}
\end{figure}

Similarly, for the $n=5$ case corresponding to $\rm Sr_2IrO_4$, the detailed spin-orbital character of the Goldstone mode and gapped mode at ${\bf q}=(0,0)$ seen in Fig. \ref{spfn5} is shown in Fig. \ref{histo_3}, explicitly illustrating the effect of extreme spin-orbital entanglement and the resulting correspondence (Fig. \ref{paneln5}) between magnetic ordering directions, spin moments, and orbital moments.   

\end{document}